\documentclass[iop]{emulateapj}
\usepackage{amsmath}

\shorttitle{HAT-P-2b Atmospheric Circulation}
\shortauthors{Lewis et al.}

\begin{document}

\title{Atmospheric Circulation of Eccentric Hot Jupiter HAT-P-2b}

\author{
Nikole K. Lewis\altaffilmark{1,6},  Adam P. Showman\altaffilmark{2}, Jonathan J. Fortney\altaffilmark{3},
 Heather A. Knutson\altaffilmark{4}, Mark S. Marley\altaffilmark{5}
}

\altaffiltext{1}{Department of Earth, Atmospheric and Planetary Sciences, Massachusetts Institute of Technology, Cambridge, MA 02139, USA;
\email{nklewis@mit.edu}}
\altaffiltext{2}{Department of Planetary Sciences and Lunar and Planetary Laboratory, 
The University of Arizona, Tucson, AZ 85721, USA}
\altaffiltext{3}{Department of Astronomy \& Astrophysics, University of California, Santa Cruz, CA 95064, USA}
\altaffiltext{4}{Division of Geological and Planetary Sciences, California Institute of Technology, Pasadena, CA 91125, USA} 
\altaffiltext{5}{NASA Ames Research Center 245-3, Moffett Field, CA 94035, USA}
\altaffiltext{6}{Sagan Postdoctoral Fellow}

\begin{abstract}
The hot-Jupiter HAT-P-2b has become a prime target for {\it Spitzer Space Telescope} observations aimed at understanding the atmospheric response 
of exoplanets on highly eccentric orbits.  Here we present a suite of three-dimensional atmospheric circulation models for HAT-P-2b that 
investigate the effects of assumed atmospheric composition and rotation rate on global scale winds and thermal patterns.  We compare and contrast 
atmospheric models for HAT-P-2b, which assume one and five times solar metallicity, both with and without TiO/VO as atmospheric constituents.  
Additionally we compare models that assume a rotation period of half, one, and two times the nominal pseudo-synchronous rotation period.  
We find that changes in assumed atmospheric 
metallicity and rotation rate do not significantly affect model predictions of the planetary flux as a function of orbital phase.  However, models in which 
TiO/VO are present in the atmosphere develop a transient temperature inversion between the transit and secondary eclipse events that results in significant 
variations in the timing and magnitude of the peak of the planetary flux compared with models in which TiO/VO are omitted from the opacity tables.  We find that 
no one single atmospheric model can reproduce the recently observed full orbit phase curves at 3.6, 4.5 and 8.0~$\mu$m, which is likely due to a 
chemical process not captured by our current atmospheric models for HAT-P-2b.  Further modeling and observational efforts focused on understanding the 
chemistry of HAT-P-2b's atmosphere are needed and could provide key insights into the interplay between radiative, dynamical, and chemical processes in 
a wide range of exoplanet atmospheres.
\end{abstract}

\keywords{atmospheric effects - methods: numerical - planets and satellites: general - planets and satellites: individual (HAT-P-2b)}

\section{Introduction}

HAT-P-2b (aka HD~147506b, $M_p=8~M_J$, $R_p=1~R_J$) was the among the first transiting extrasolar planets 
discovered to have a significant orbital eccentricity ($e\sim0.5$) \citep{bak07a}.  
Previous studies of the highly eccentric HD~80606b ($e\sim0.9$) focused on orbital phases near the planet's periapse passage and secondary 
eclipse (when the planet passes behind its host star) to make the first determination of the radiative timescale at infrared wavelengths of an 
exoplanet atmosphere \citep{lau09}.  However, the long duration of HD~80606b's orbit (111~days) makes it nearly impossible 
to observe the planet through the entirety of its orbit.
HAT-P-2b has an orbital period just over 5.6 days \citep{bak07a, win07, loe08, pal10}, which
makes it an advantageous target for full-orbit observations to study the evolution of planetary flux with orbital phase.

The incident flux on HAT-P-2b from its stellar host at periapse is ten times that at apoapse, which should cause 
large variations in atmospheric temperature, wind speeds, and chemistry.  
We recently used full orbit {\it Spitzer} 
observations at 3.6, 4.5, and 8.0~$\mu$m of the HAT-P-2 system to measure HAT-P-2b's flux variations as a function 
of orbital phase and constrain the radiative timescale on the planet near periapse to be between two and eight hours \citep{lew13}.  Here 
we present three-dimensional atmospheric models for HAT-P-2b with a range of atmospheric compositions and 
rotation rates that can be compared directly to these observations.  These atmospheric models for HAT-P-2b add to 
the growing body of work investigating the atmospheric dynamics of close-in Jovian sized planets, most of which 
focus on planets with circular orbits 
\citep[e.g.][]{sho02, coo05, cho08, sho09, lan07, dob10, men09, rau10, heng11, per12, heng12, sho13, rau13}.  

Only a handful of studies have investigated the atmospheric circulation of eccentric `hot Jupiters' \citep{lan08,lew10,kat13}.  
The simulations of HAT-P-2b presented here provide theoretical predictions for the expected variations in the 
planetary flux as a function of orbital phase under a range of model assumptions.  These theoretical predictions can be compared 
directly with observational data to reveal the complex radiative, dynamical, and 
chemical processes in HAT-P-2b's atmosphere that may or may not be fully captured by current state-of-the-art general circulation models.  
Detailed study of planets subject to strong time-variable forcing such as HAT-P-2b present a unique opportunity to understand the competing effects of radiation, dynamics, and 
chemistry that shape the global circulation patterns of extrasolar planets.

In the following sections we first describe the set-up of our three-dimensional atmospheric model for HAT-P-2b (\S\ref{hat2_model}) 
including variations in the assumed atmospheric chemistry and rotation period for the planet.  We present the results from 
our model simulations in \S\ref{hat2_results} including the thermal structure and winds that develop in each case and the 
resulting theoretical light curves and spectra.  In \S\ref{hat2_discussion} we discuss the overall trends we see in our simulations 
of HAT-P-2b's atmosphere and compare and contrast our model predictions with what has been observed.  Finally, in \S\ref{hat2_conclusions}
we present a summary of our results and discuss future work.

\section{Atmospheric Model}\label{hat2_model}

We employ a three-dimensional coupled atmospheric radiative transfer and dynamics model, the
Substellar and Planetary Atmospheric Radiation and Circulation (SPARC) model,   
to investigate the atmospheric circulation of HAT-P-2b.  The SPARC model was developed specifically to explore 
atmospheric circulation on exoplanets and is described 
in detail in \citet{sho09} as applied to HD~189733b and HD~209458b,  \citet{lew10} as applied to GJ~436b, 
and \citet{kat13} as applied to an HD~189733b-like planet on a range of eccentric orbits.  
The SPARC model employs the MITgcm \citep{adc04} to treat the atmospheric dynamics and utilizes  
a two-stream variant of the non-gray radiative transfer model of \citet{mar99} in order to realistically determine the magnitude of 
radiative heating/cooling at each grid point.  The simulations 
presented here take advantage of the cubed-sphere grid \citep{adc04} at a resolution of C16 (roughly $32\times64$ in latitude and longitude) to 
solve the relevant dynamic and energy equations.   The vertical dimension in these simulations spans the pressure ($p$) range from 1000~bar 
to 0.2~mbar with 45 vertical levels, evenly spaced in $\log(p)$, and a top layer that extends to zero pressure.

Atmospheric opacities are computed according to \citet{fre08} as a function of pressure, temperature, 
and wavelength assuming a range of atmospheric compositions in thermochemical equilibrium \citep{lod02,lod06}. 
We then divide the full wavelength range into 11 wavelength/frequency bins 
for greater computational efficiency \citep[see][for a full discussion of the 11 wavelength bin version of SPARC]{kat13}.  
Within each wavelength bin we utilize the correlated-$k$ method \citep[e.g.][]{mla97, goo89, fu92, mar99} to 
determine the overall opacity using eight gauss points.  This results in 88 separate radiative transfer 
calculations to determine upward and downward fluxes for a single grid point.  It is important to note that 
this approach is far more accurate than simple band models \citep[e.g.][]{dob12} and statistically 
incorporates information from several million wavelength points in each of our wavelength bins.

In this study, we consider atmospheric compositions of both one times ($1\times$) and five times ($5\times$) 
solar metallicity.  In the $5\times$ solar metallicity case, the abundance of all elements besides hydrogen and helium are enhanced 
by five times standard solar values.  Given HAT-P-2b's mass and radius it is possible that it is enriched in heavy elements \citep[e.g.][]{for08b, gui05}.
We also consider atmospheric compositions with and without TiO and VO.  As proposed in \citet{for08a} \citep[see also,][]{hub03}, the presence of the strong 
optical absorbers TiO and VO in a planet's atmosphere could be responsible for atmospheric temperature inversions that have 
been observed for a number of extrasolar planets.  Here TiO and VO are assumed to be at their equilibrium abundances, depending on the local temperature 
and pressure, following \citet{lod02b}.  The computed equilibrium abundances and corresponding opacities do account for rainout of condensed species (including TiO and VO).  
Although we do not rigorously treat possible `cold-trapping' \citep{hub03}, our opacity tables in which TiO and VO have been
excluded represent a case where these chemical species have been cold
trapped in the deep atmosphere (below 100 bars) or condensed out on the 
nightside of the planet (day/night cold trap) as suggested by the models of 
HD~209458b by \citet{par13} and the observations of WASP-12b by \citet{sing2013}.
Certainly it is possible for cold-trapping to be incomplete, leaving 
TiO and VO at a few percent of their equilibrium abundance levels, but here we only consider the 
end-member state of complete TiO/VO removal.

\begin{figure}
\centering
  \includegraphics[width=0.5\textwidth]{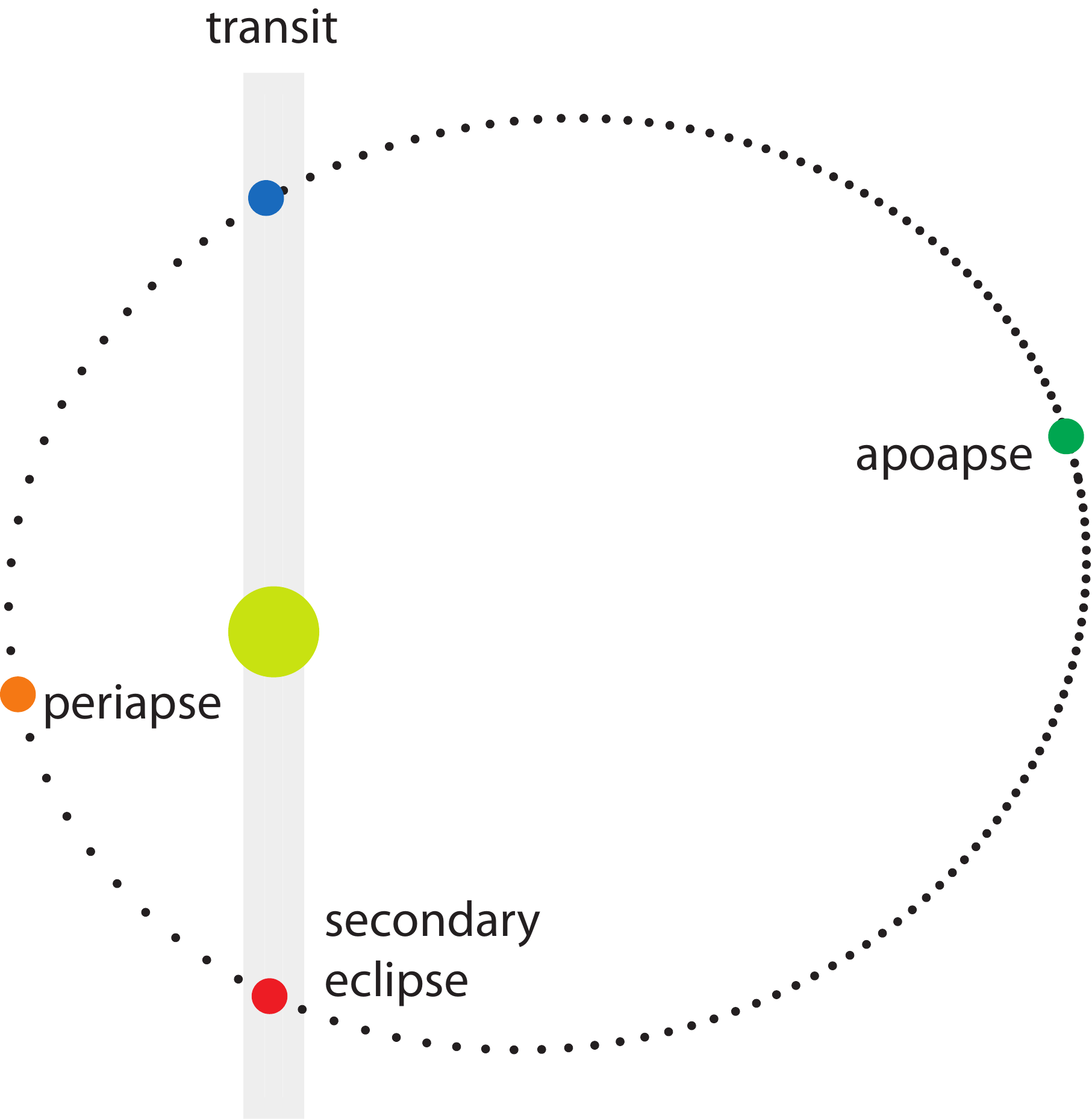}
  \caption{Orbit of HAT-P-2b assuming parameters from \citet{lew13}.  Dots along the orbital path represent points where data was extracted 
  to produce Figures \ref{hat2_model_lc}, \ref{hat2_model_flux}, and \label{hat2_pt_co_ch4}.  The temporal interval between the points 
  is one hour.  Orbital motion is 
  in the counterclockwise direction.  Colored dots represent points near periapse (orange), transit (blue), apoapse (green), and 
 secondary eclipse (red), which correspond to the colored spectra and pressure temperature profiles presented in Figures~\ref{hat2_model_flux} 
  and \label{hat2_pt_co_ch4} respectively.}\label{hat2_orb_fig}
\end{figure}

\begin{figure*}
\centering
 \includegraphics[width=0.45\textwidth]{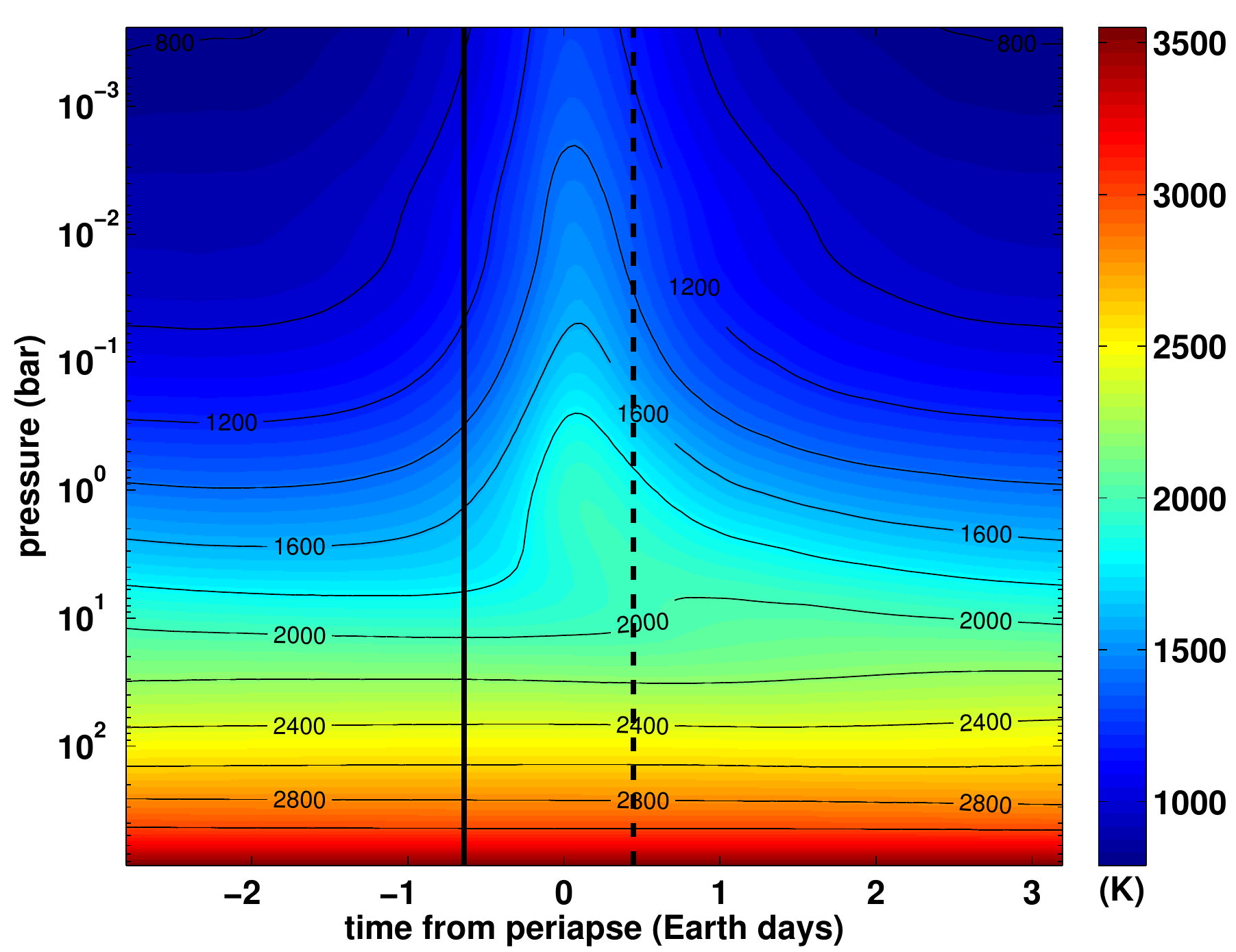}
 \includegraphics[width=0.45\textwidth]{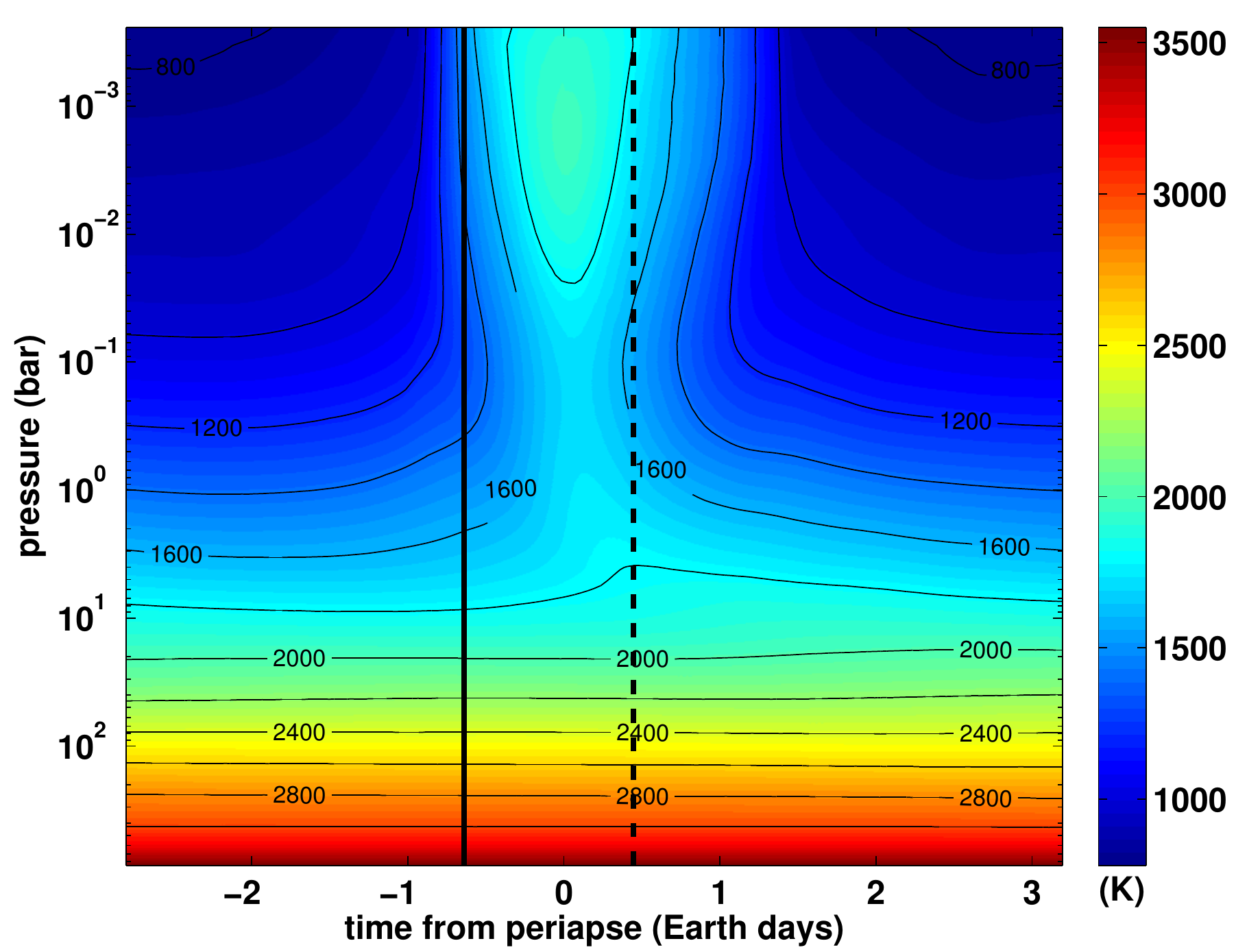}\\
 \includegraphics[width=0.45\textwidth]{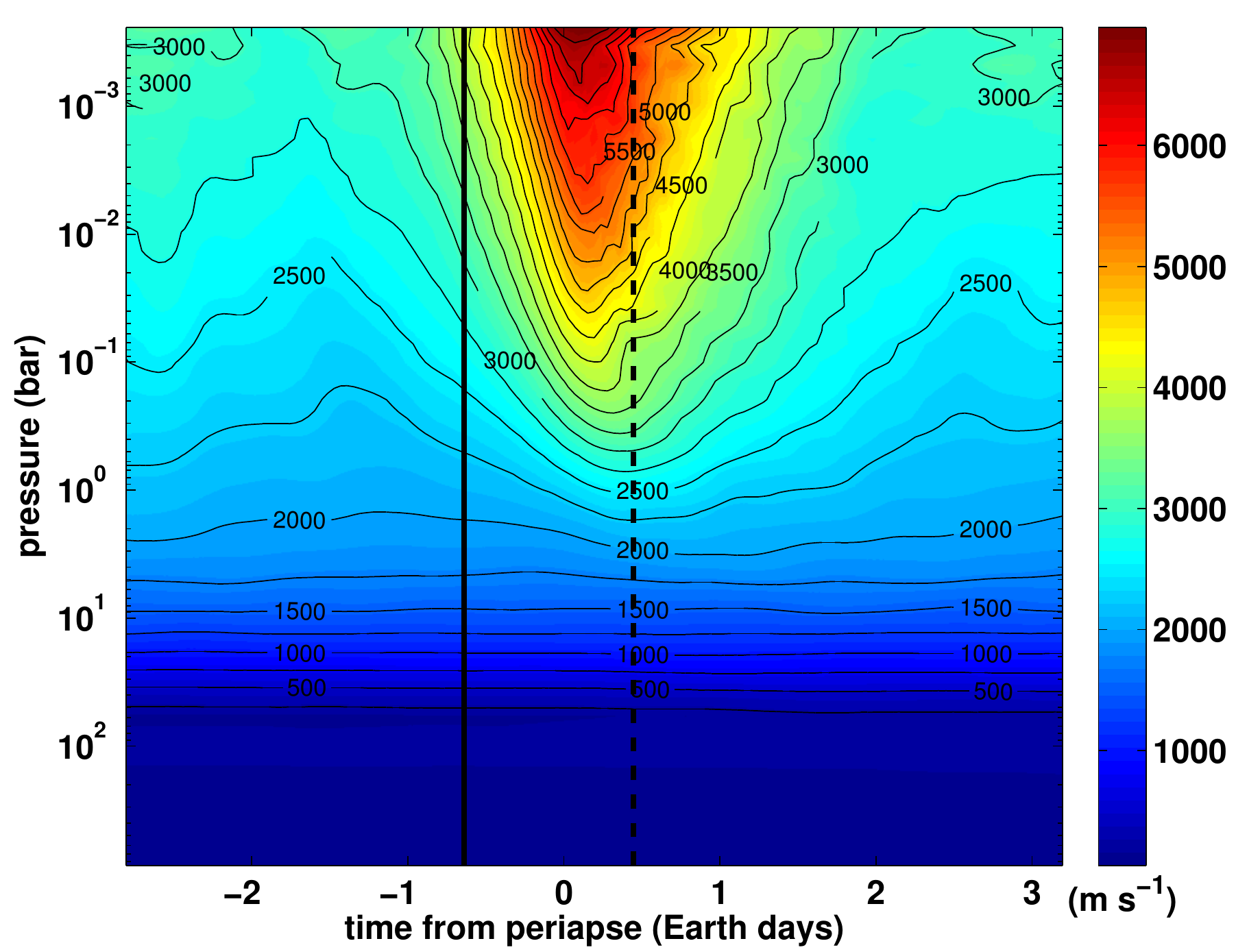}
 \includegraphics[width=0.45\textwidth]{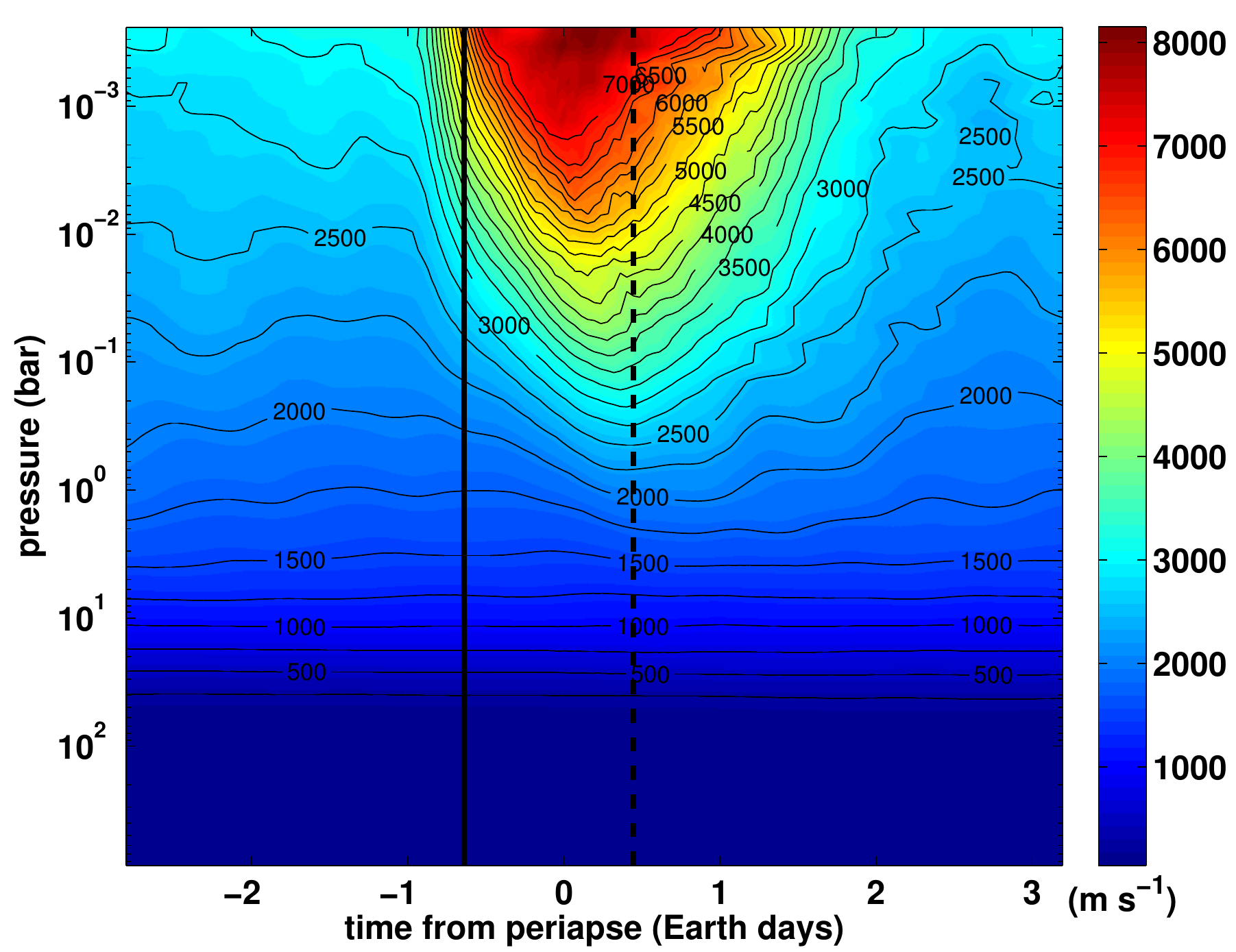}\\
  \caption{Horizontally averaged temperature (top) and RMS horizontal velocity (bottom) as a function of time 
relative to periapse passage for our simulations without TiO/VO (left) and with TiO/VO (right). 
The temperatures represent averages over latitude and longitude as a 
function of pressure.  RMS horizontal velocities are calculated according to Equation~\ref{vrms}.  
The times of transit and secondary eclipse are represented by the vertical solid and dashed lines respectively.  
 }\label{hat2_vrms_vs_time}
\end{figure*}

In our model we assume the planetary and stellar parameters for the HAT-P-2 system given in \citet{lew13}.   We assume a nominal rotation period
of 1.9462563~days for HAT-P-2b, which we calculate using the pseudo-synchronous rotation relationship presented in \citet{hut81}.
Additionally, we construct models with a rotation periods twice ($2\times$) and half ($0.5\times$) the nominal pseudo-synchronous rotation period
for our standard 1$\times$ solar composition model that includes TiO and VO. 
We initialize the model with wind speeds set to zero everywhere and each column of the grid assigned the same pressure-temperature 
profile.  This initial pressure-temperature profile is derived from one-dimensional radiative-equilibrium calculations described in \citet{for05,for08a}
assuming no atmospheric dynamics and that HAT-P-2b is at the periapse of its orbit with an intrinsic effective temperature of 300~K.  We find no difference in the equilibrated state of models initiated near 
apoapse with a one-dimensional pressure-temperature profile assuming the planet's apoapse distance as opposed to periapse with 
a one-dimensional pressure-temperature profile assuming the planet's periapse distance.  This is in line with the results 
from \citet{liu13}, which found that the steady-state solution of hot Jupiter circulation models 
are insensitive to the initial conditions.  The time-varying distance of the planet with 
respect to its host star, $r(t)$, is determined using Kepler's equation \citep{mur99} and used to update the incident flux on the planet 
at each radiative timestep.  A diagram of HAT-P-2b's orbit is presented in Figure~\ref{hat2_orb_fig}.

We solve for the relevant dynamic and energy equations using a dynamical timestep of 5~s. 
For computational efficiency, the radiative timestep used to update the radiative fluxes is varied as a function of orbital 
phase.  We use a radiative timestep of 5~s, 25~s, and 50~s when the planet is near the periapse, midpoint, and apoapse 
of its orbit respectively.   We apply a fourth-order Shapiro filter in the horizontal direction to both velocity components and 
the potential temperature over a timescale equivalent to the dynamical timestep in order to reduce small scale grid noise 
while minimally affecting the physical structure of the wind and temperature fields at the large scale.  We integrate our model
until the orbit averaged root mean square (rms) velocity given by 
\begin{equation}
V_{rms}(p)=\sqrt{\frac{\int (u^2+v^2)\,dA}{A}}\label{vrms}
\end{equation}
where the integral is a global (horizontal) integral over the globe, $A$ is the horizontal area of the globe, 
$u$ is the east-west wind speed, and $v$ is the north-south wind speed, reaches a stable configuration.
Any further increases in wind speeds will be small and confined to pressures well 
below the mean photosphere so as not to affect any synthetic observations derived from our model atmosphere.
For the simulations presented here the nominal integration time is $\sim$300 simulated days, which is equivalent to 
more than 50 planetary orbits.    
We find that the total angular momentum in our simulations varies by less than 0.1\% and is hence well conserved.

\section{Results}\label{hat2_results}

\subsection{Thermal Structure and Winds}

\begin{figure*}
\centering
 \includegraphics[width=0.45\textwidth]{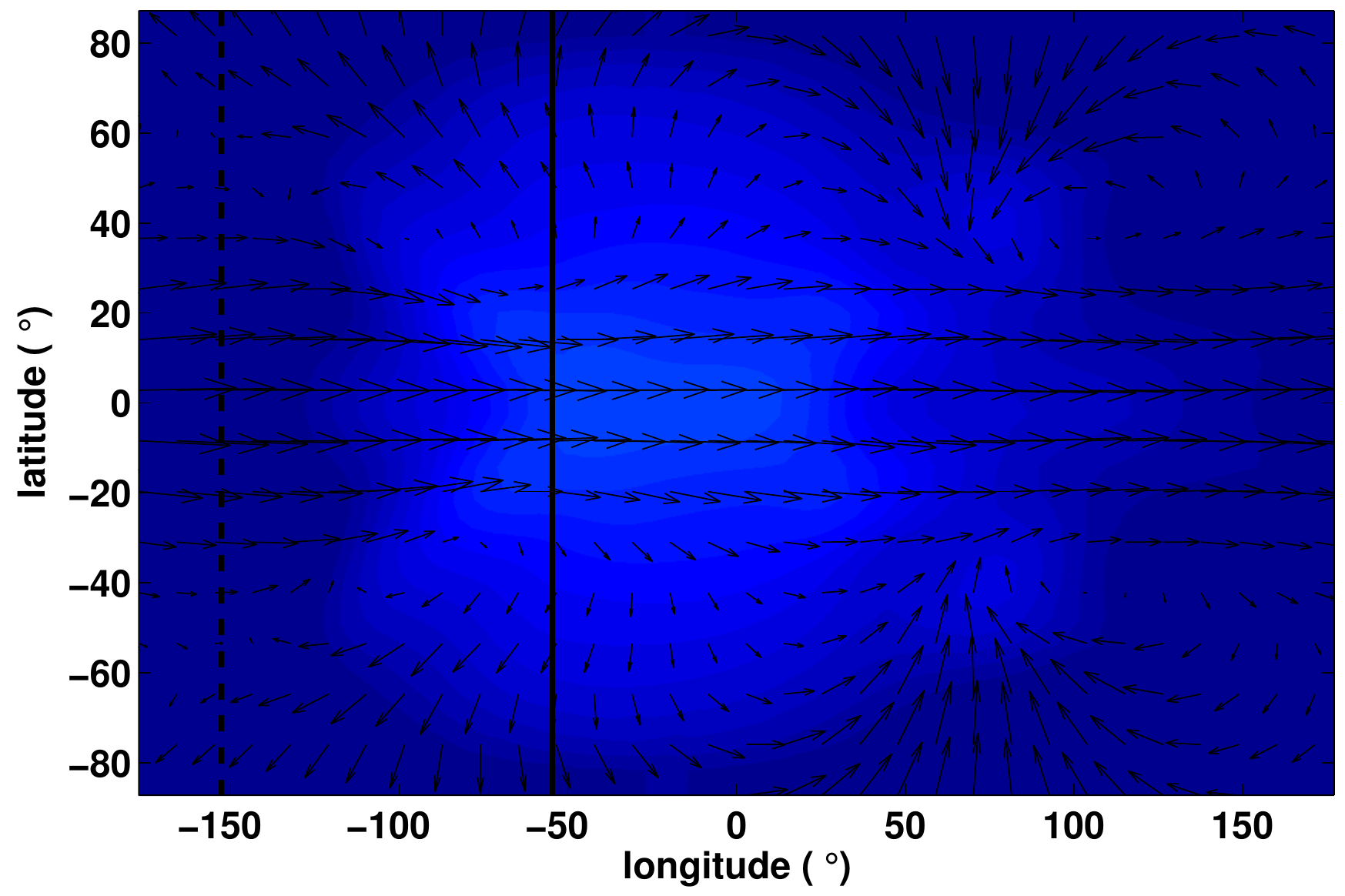}
 \includegraphics[width=0.45\textwidth]{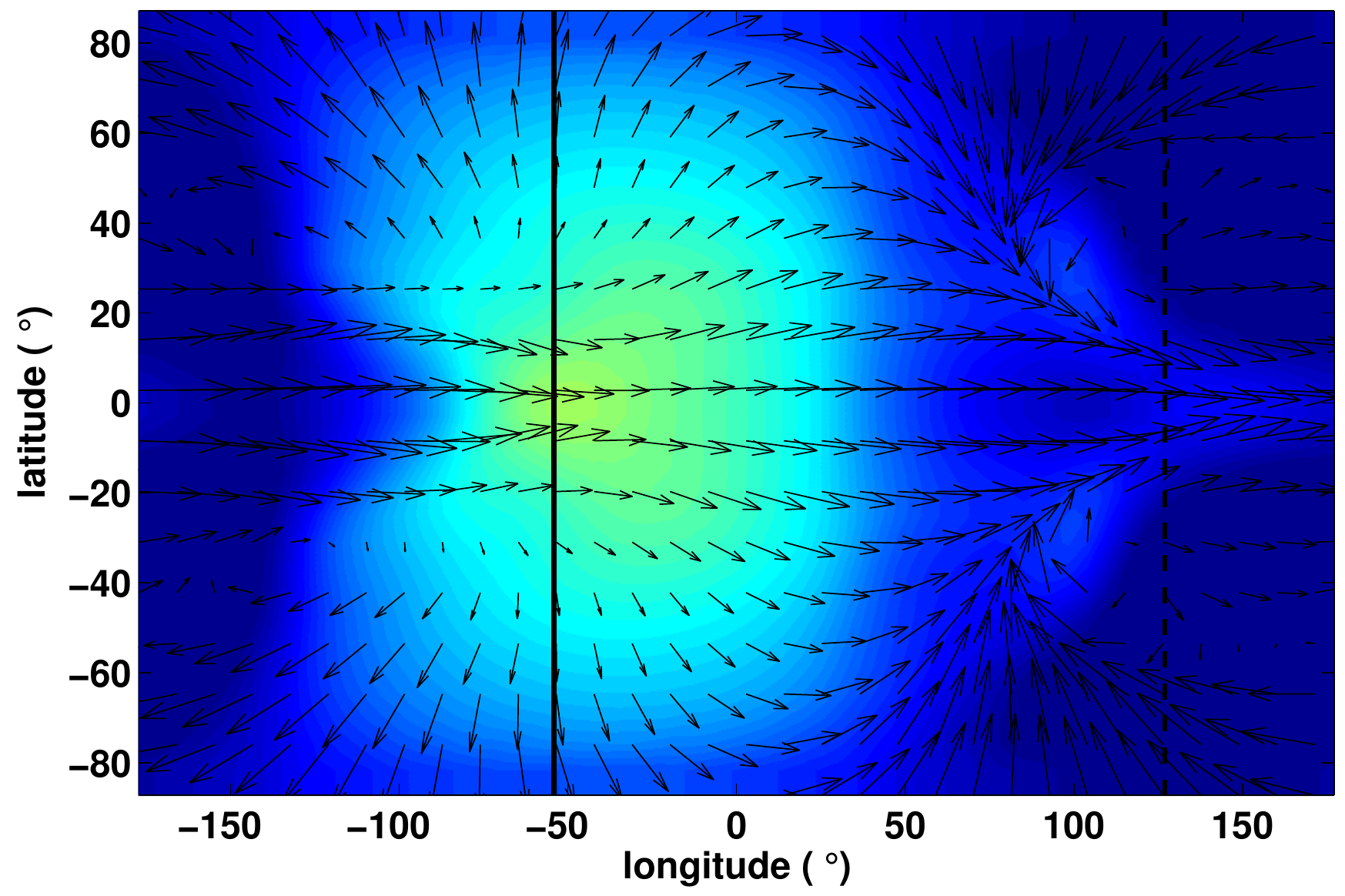}\\
 \includegraphics[width=0.45\textwidth]{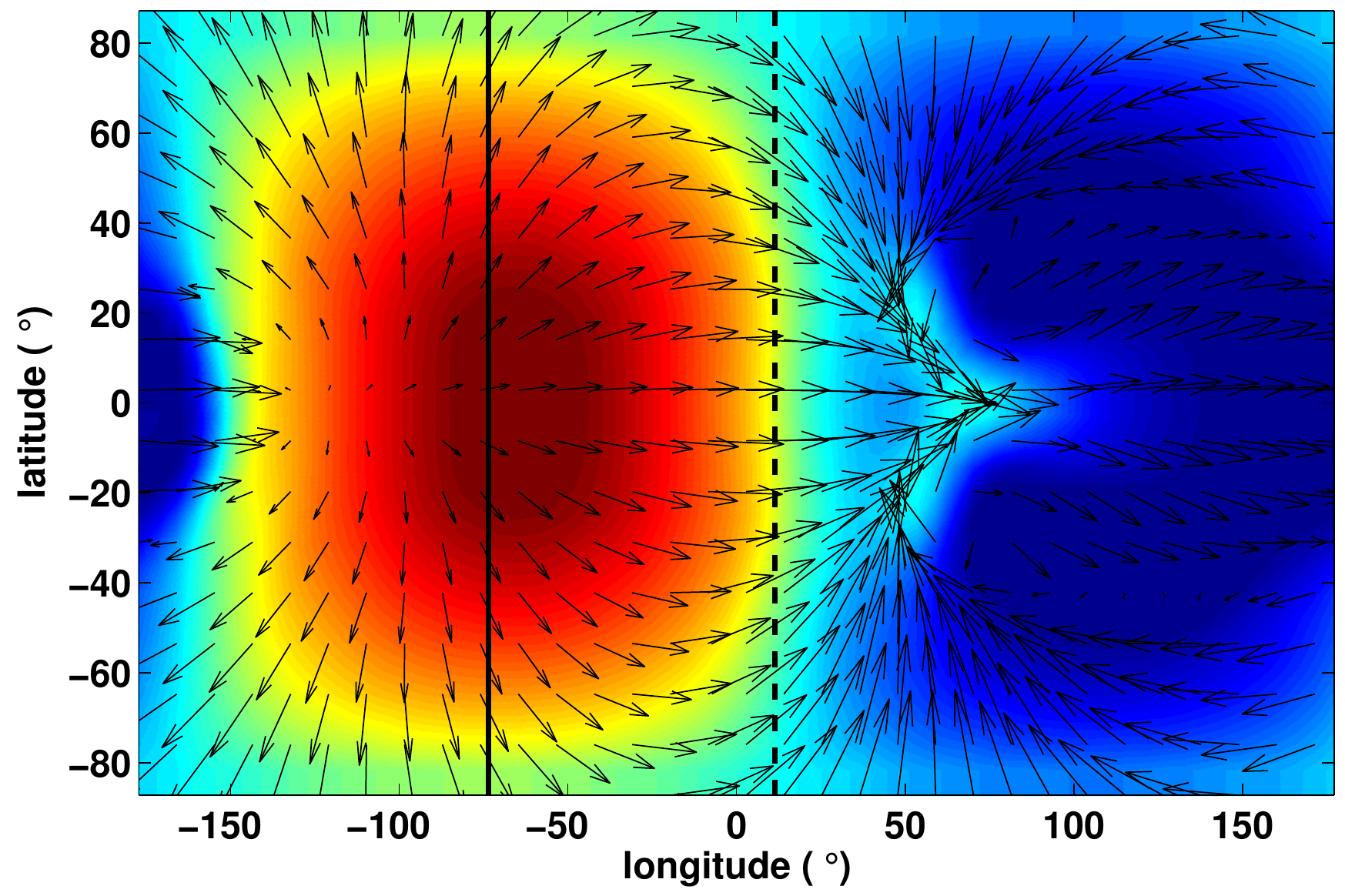}
 \includegraphics[width=0.45\textwidth]{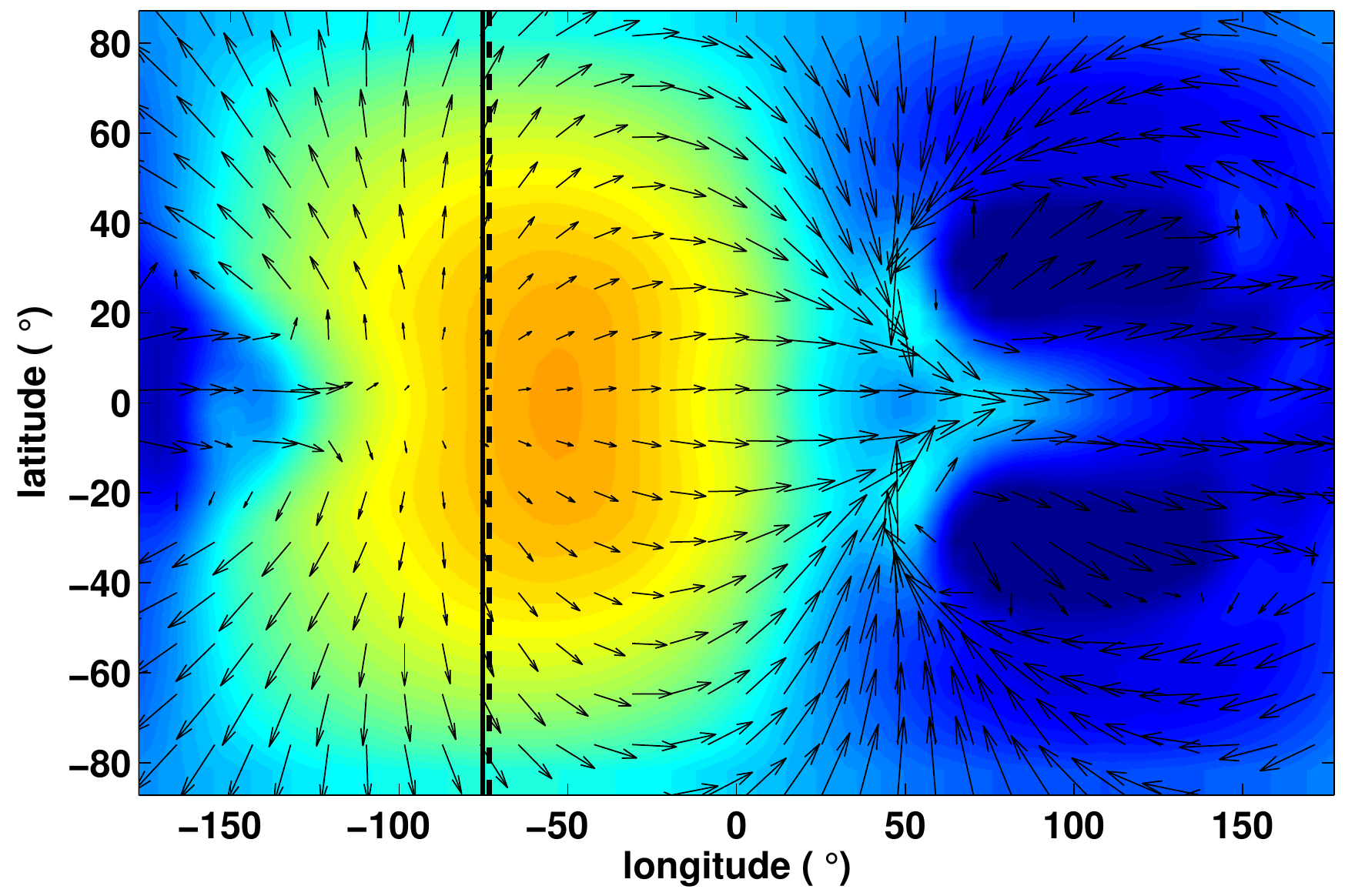}\\
 \includegraphics[width=0.60\textwidth]{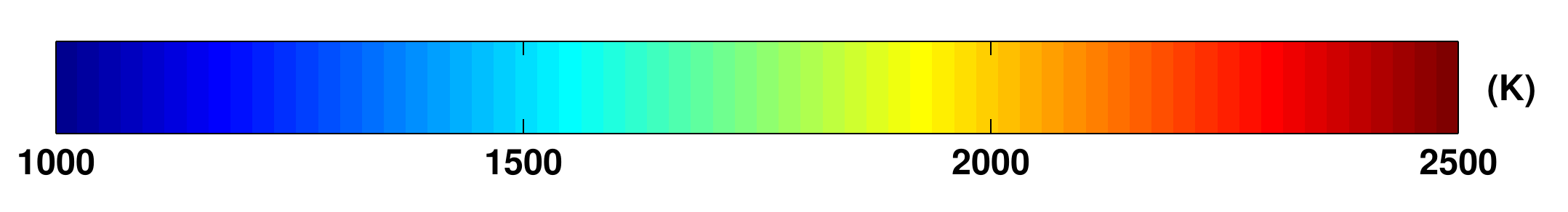}
  \caption{Temperature (colorscale) and winds (arrows) at the 100~mbar level of our 1$\times$ solar model without TiO/VO for time near apoapse (top left), 
  transit (top right), periapse (bottom left) and secondary eclipse (bottom right).  The longitude of the substellar and sub-earth points are 
  indicated in each panel by the solid and dashed vertical lines respectively.  A substantial day/night temperature contrast exist for the majority 
  of the HAT-P-2b's orbit in our models.  The offset between the substellar longitude and the peak in the planet's temperature varies between 
  $\sim 10^{\circ}$ and $\sim 50^{\circ}$ during the planet's orbit.}\label{hat2_temp_100mbar_noTiO}
\end{figure*} 
\begin{figure*}
\centering
 \includegraphics[width=0.45\textwidth]{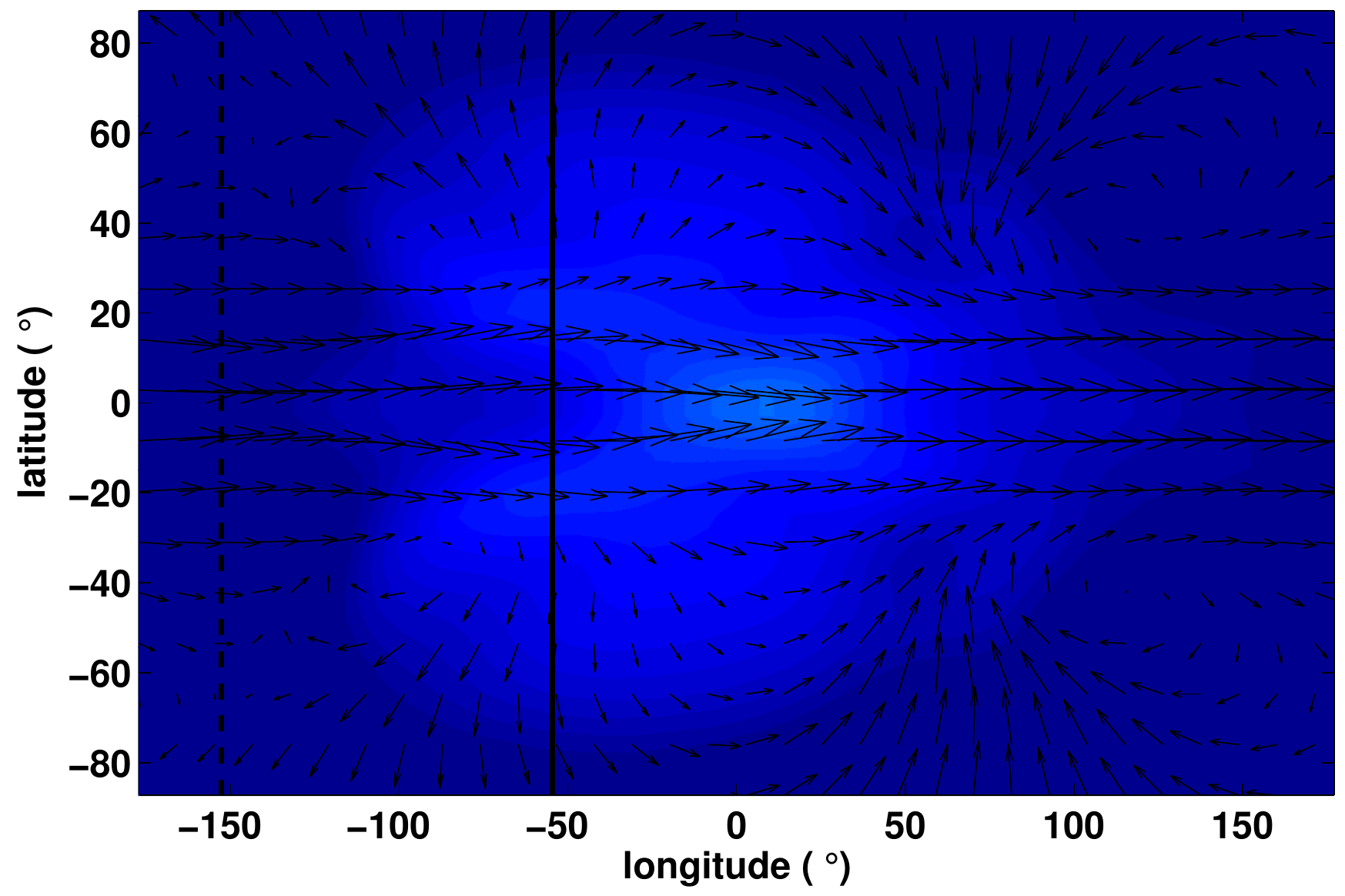}
 \includegraphics[width=0.45\textwidth]{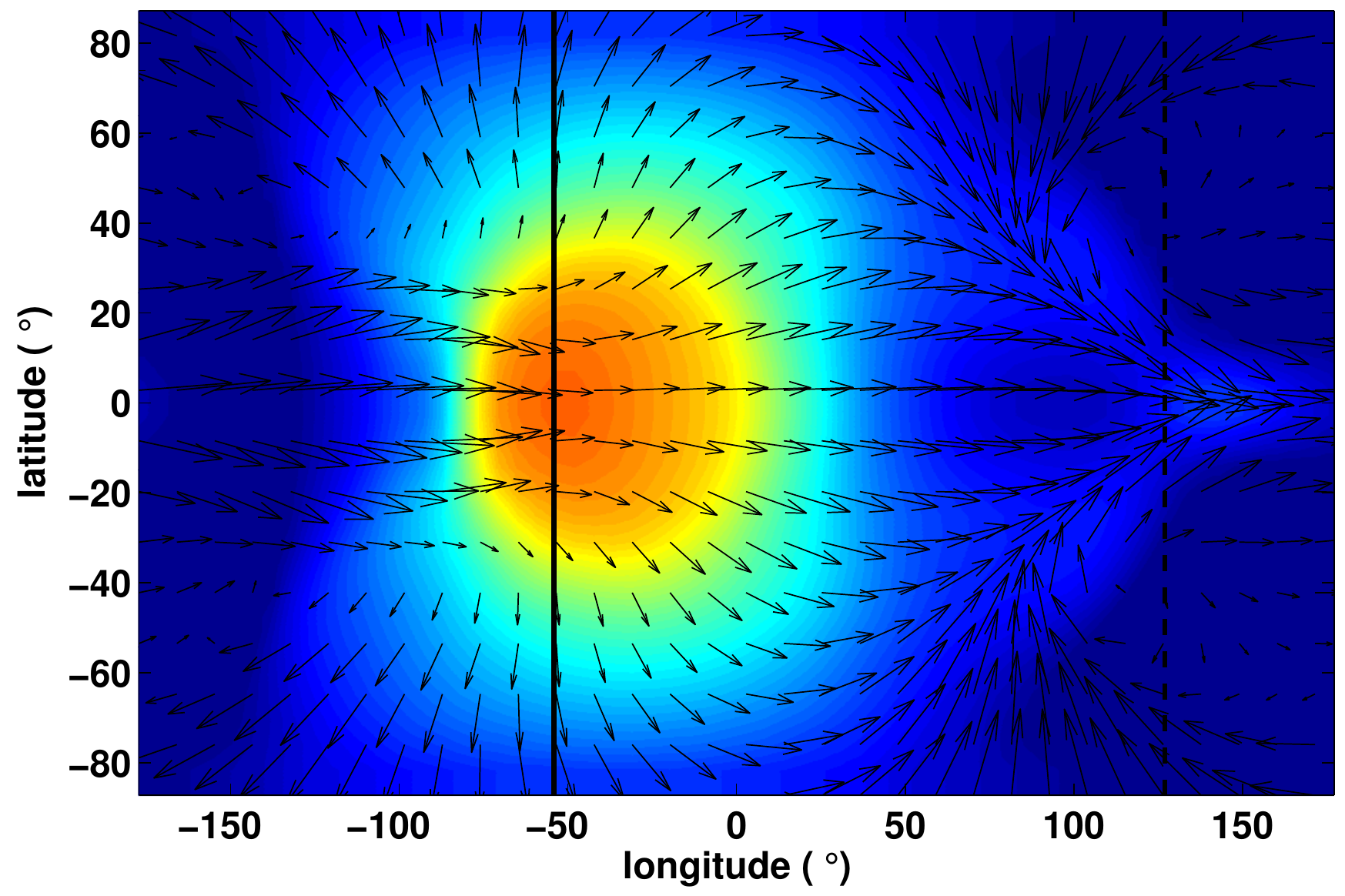}\\
 \includegraphics[width=0.45\textwidth]{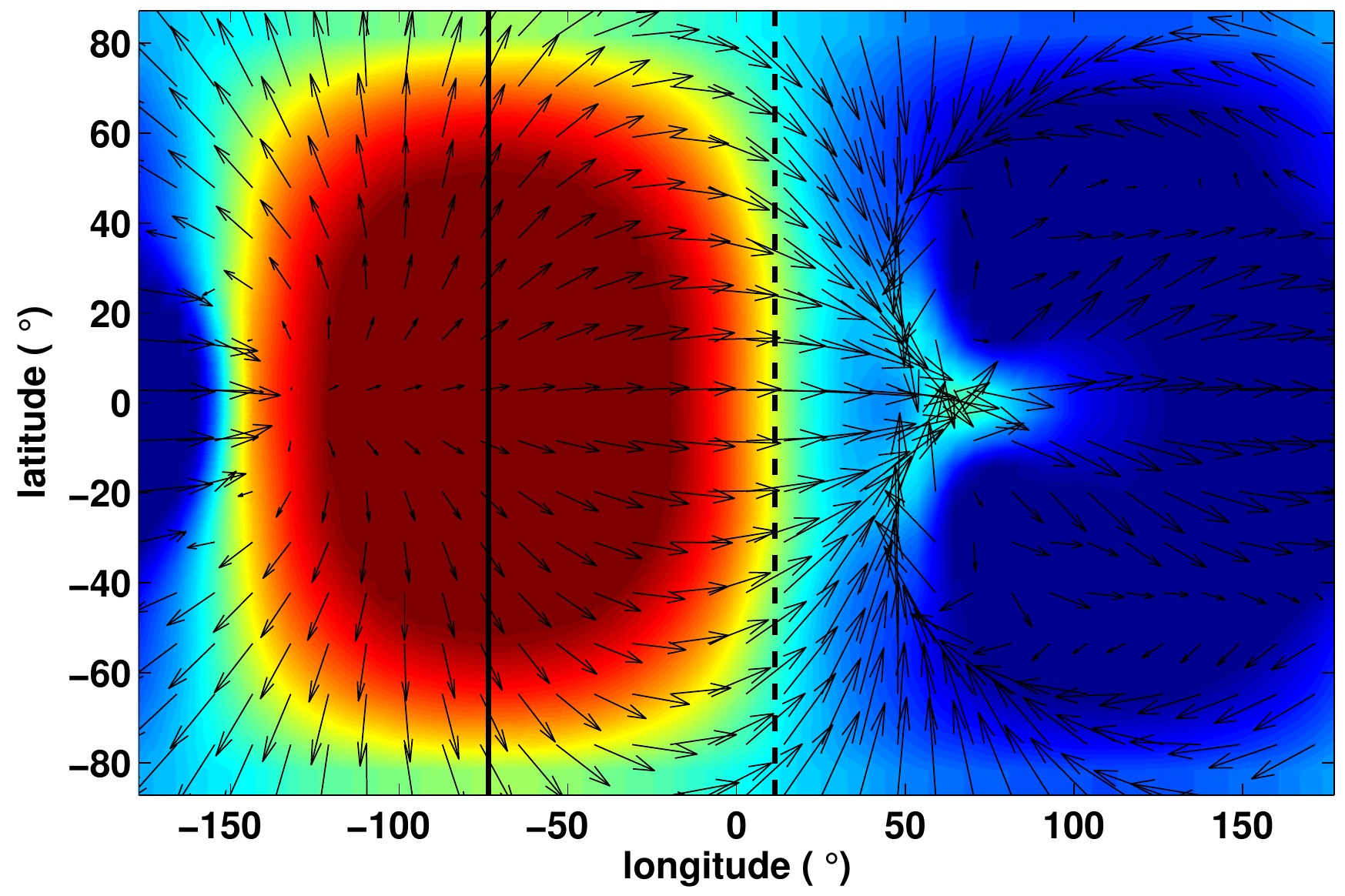}
 \includegraphics[width=0.45\textwidth]{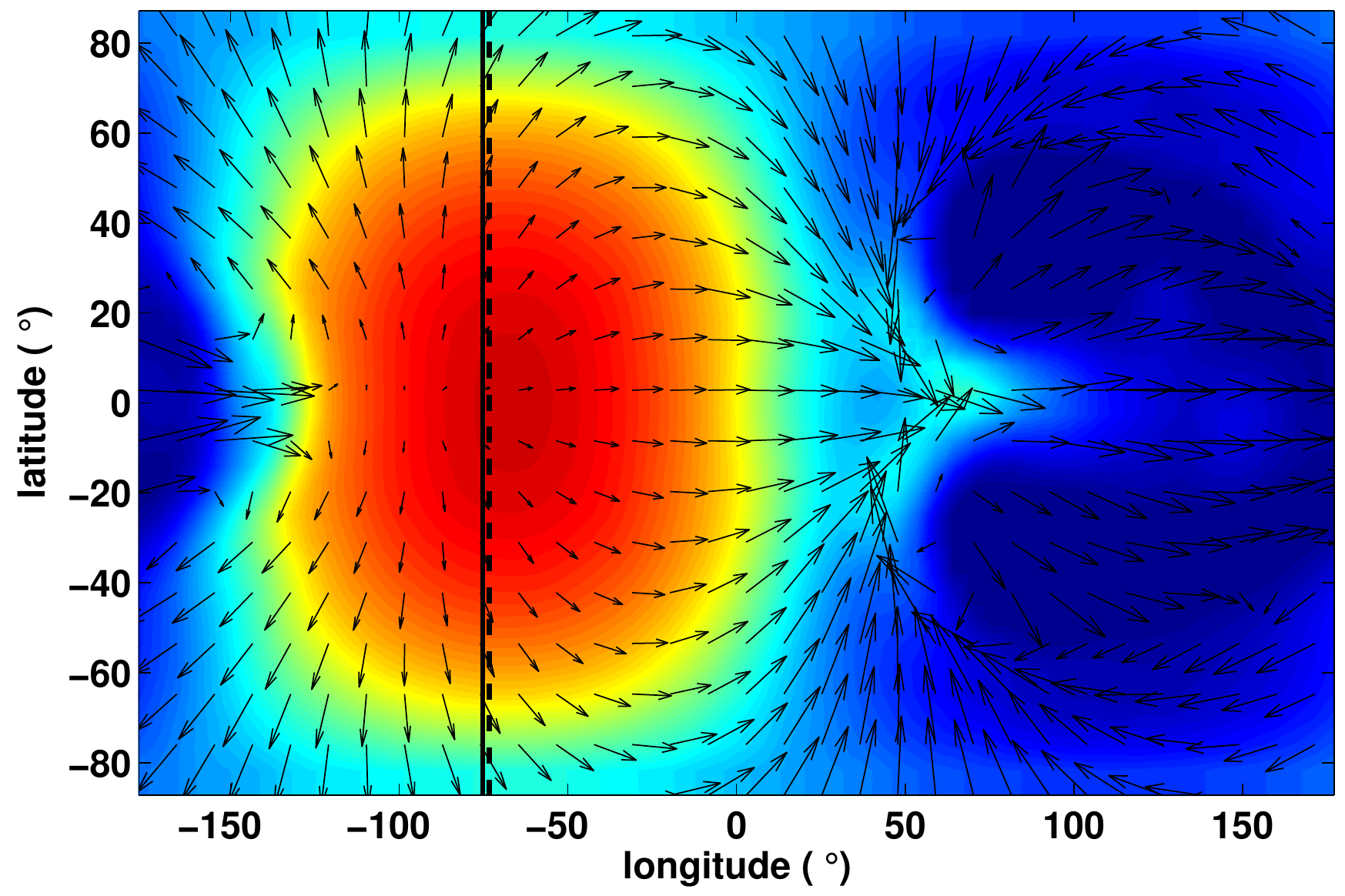}\\
 \includegraphics[width=0.60\textwidth]{colorbar_1000_2500.pdf}
  \caption{Temperature (colorscale) and winds (arrows) at the 100~mbar level of our 1$\times$ solar model with TiO/VO for time near apoapse (top left), 
  transit (top right), periapse (bottom left) and secondary eclipse (bottom right).  The longitude of the substellar and sub-earth points are 
  indicated in each panel by the solid and dashed vertical lines respectively.  A substantial day/night temperature contrast exist for the majority 
  of the HAT-P-2b's orbit in our models.  The offset between the substellar longitude and the peak in the planet's temperature varies between 
  $\sim 0^{\circ}$ and $\sim 20^{\circ}$ during the planet's orbit.}\label{hat2_temp_100mbar_eqchem}
\end{figure*}

Our atmospheric models of HAT-P-2b exhibit large variations in both temperature and wind speeds as a function 
of orbital phase.  Figure~\ref{hat2_vrms_vs_time} presents temperature, averaged over latitude and longitude, and RMS velocity as 
a function of both pressure and simulated time for a single orbit of HAT-P-2b for our 1$\times$ solar models both with and without TiO/VO.  
Below the 10~bar level, average temperatures and wind speeds in our models are not strongly affected by the changes in incident flux on HAT-P-2b during its orbit 
and are similar between models with and without TiO/VO as an atmospheric constituent.  
Above the 1~bar level, both average temperature and wind speeds are a strong function of orbital phase and display marked difference between models that include 
TiO/VO and models in which TiO/VO has been excluded.  This difference in global temperature and wind speeds in models with TiO/VO as compared to models without 
TiO/VO is the result of the development of a transient thermal inversion as seen in the top right panel of Figure~\ref{hat2_vrms_vs_time}.

The timing of the peak of global wind speeds and temperatures also varies between models that include TiO/VO and those that do not.  
We find that peak global temperatures above the 10~bar level occur $\sim$1~hours and $\sim$2~hour after periapse 
for the models with and without TiO/VO respectively.  A similar variation in the timing of peak wind speeds above 1~bar is also seen with 
the peak occurring $\sim$3~hours after periapse for models with TiO/VO and $\sim$4~hours after periapse for models that 
do not include TiO/VO.    It is also interesting to note that elevated wind speeds near periapse are much larger in magnitude and persist for a significantly longer time in our models 
with TiO/VO as compared to those models with out TiO/VO.  

It is also important to note the timescales that we expect the increase and decrease in the planetary flux to 
occur over and how those timescales vary with pressure as they will shape the observed phase curves.  
At all pressures, the timescales required for the planet to heat up are shorter than those required for the planet to 
cool down.  This means that we expect HAT-P-2b's temperature to remain elevated above pre-periapse levels 
well into the apoapse of its orbit and a minimum in the planetary flux to occur sometime between apoapse and 
transit.  The slope of the planetary cool-down period greatly increases with increasing pressure.  By comparing the 
cool-down timescales measured by the phase-curves at various wavelengths, rough constraints can be placed 
on the pressure levels being probed by each bandpass.  

\begin{figure*}
\centering
 \includegraphics[width=0.45\textwidth]{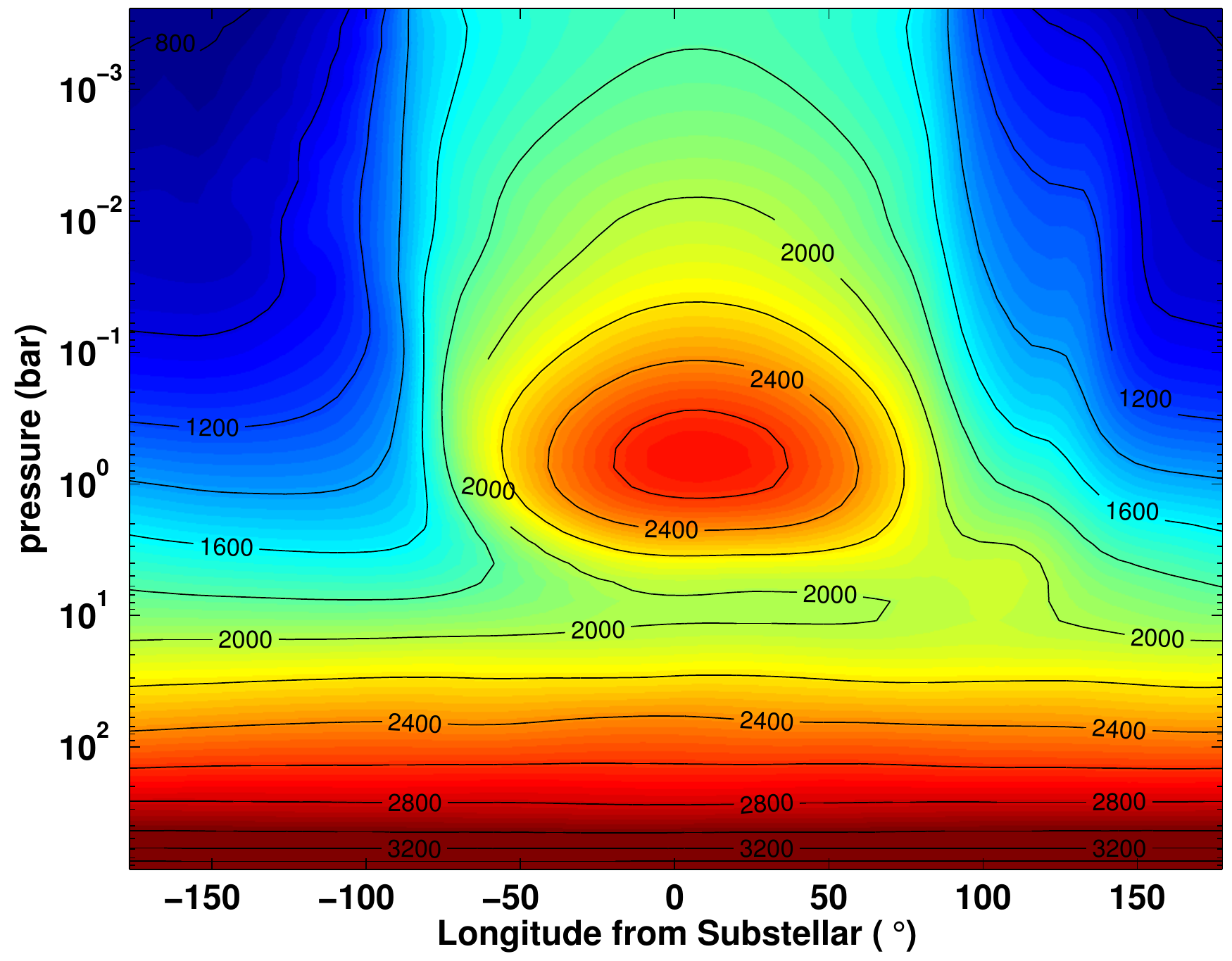}
 \includegraphics[width=0.45\textwidth]{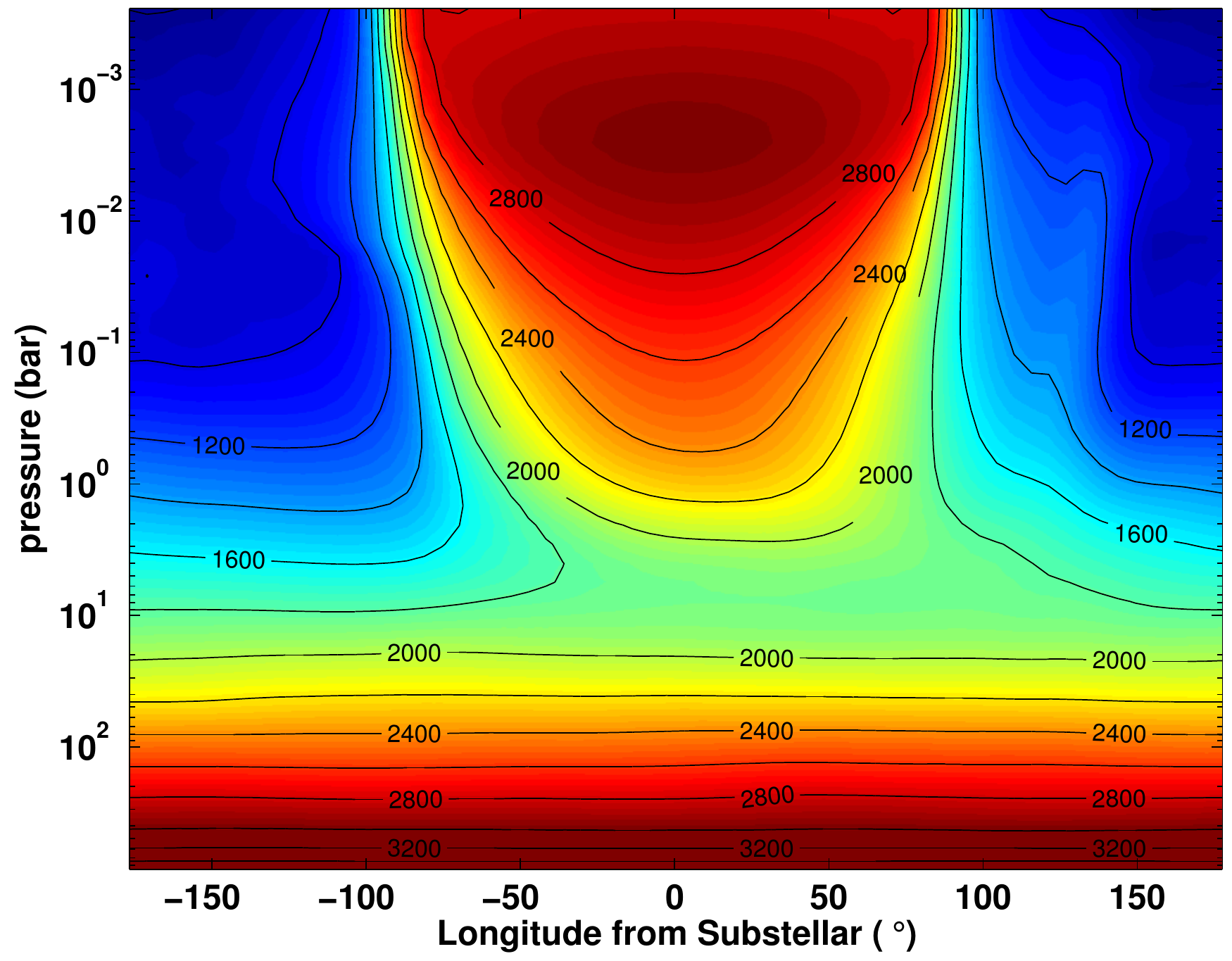}\\
 \includegraphics[width=0.45\textwidth]{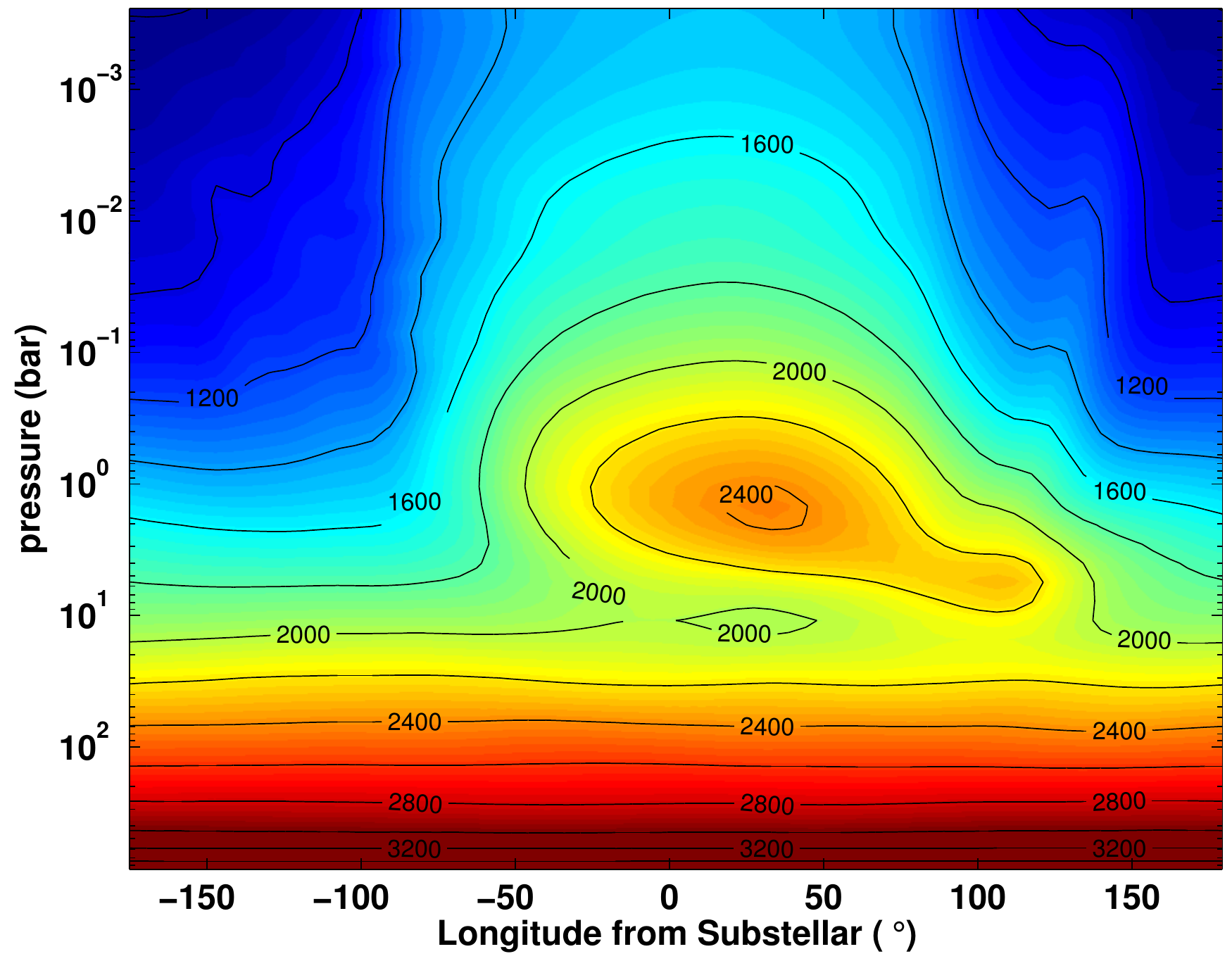}
 \includegraphics[width=0.45\textwidth]{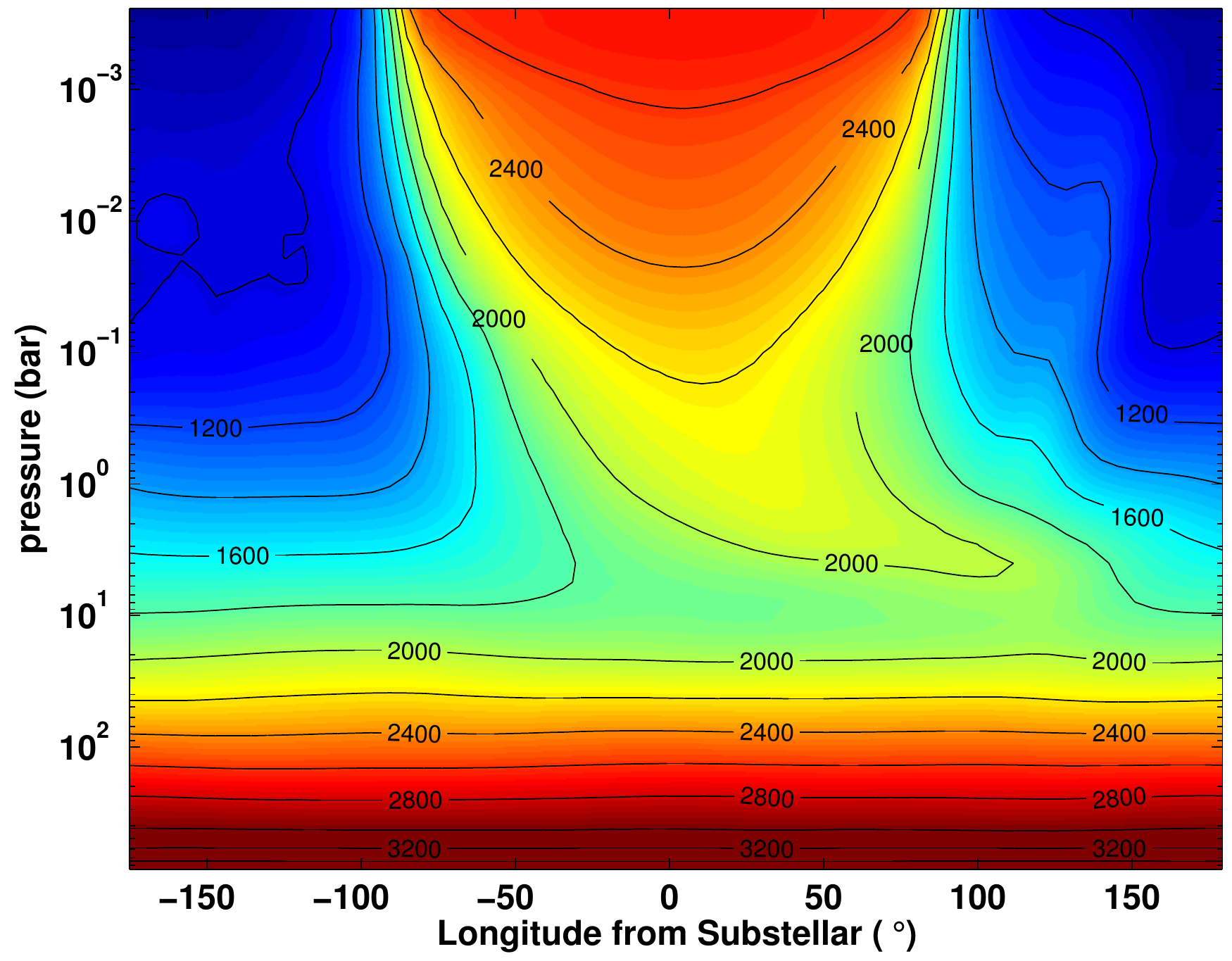}\\
 \includegraphics[width=0.60\textwidth]{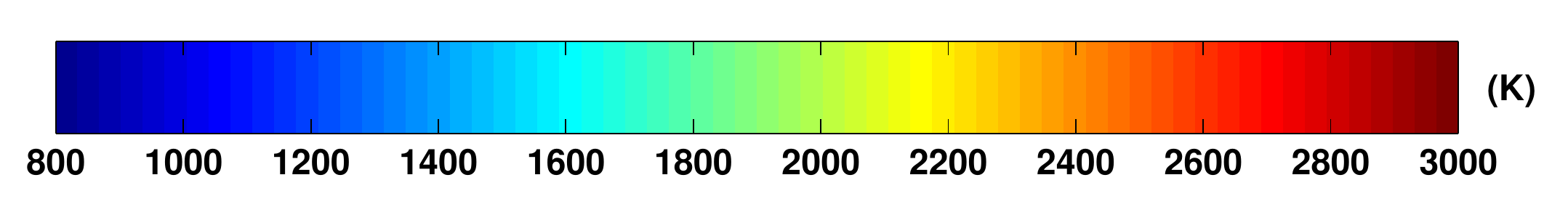}
  \caption{Temperature averaged in latitude (colorscale) as a function of pressure and degrees from the substellar longitude at 
  periapse (top) and secondary eclipse (bottom) from our models with (right) and without (left) TiO/VO.  Temperatures represent 
  average values weighted by $\cos(\phi)$, where $\phi$ is latitude.  Note how the location of the hottest point on the planet evolves 
  between the periapse and secondary eclipse events and differs significantly between models with and without TiO/VO.}\label{hat2_hot_spot}
\end{figure*} 

\subsubsection{Evolution of Thermal and Wind Patterns}\label{evolution}

In addition to changes in average temperature, it is important to consider day/night temperature contrasts that can also 
affect the observed planetary flux as a function of orbital phase.  Figures~\ref{hat2_temp_100mbar_noTiO} and  \ref{hat2_temp_100mbar_eqchem} presents slices from 
our 1$\times$ solar model without and with TiO/VO, respectively.  These slices are from the 100~mbar level of our simulations, which is near the predicted infrared 
photosphere of the planet, and represent snapshot near the apoapse, transit, periapse, and secondary eclipse events of HAT-P-2b's 
orbit.  The basic winds patterns that develop at the 100~mbar level are very similar between the simulations with and without TiO/VO, at 
all orbital phases.  The key differences between the simulations with and without TiO/VO is the magnitude of thermal gradient between 
the day and night sides of the planet that results from the dayside thermal inversion that develops in the simulations that include TiO/VO.  

We find that a significant day/night temperature contrast exists in our simulations at pressures less than 10~bar
for the majority of HAT-P-2b's orbit.  Temperature differences between the day and night side hemispheres are greatest near 
the periapse and weakest near apoapse.  This is not surprising given that changes in radiative timescales can be related 
to atmospheric temperatures ($T$) by $\tau_{rad}\propto T^{-3}$.  Even in the presence of strong winds, or other 
dynamical processes such as wave propagation, significant 
horizontal temperature contrasts will exist if $\tau_{rad} \ll \tau_{dyn}$, 
where $\tau_{dyn}$ is the relevant dynamical timescale \citep{sho10, per13}. 
It is important to note that $\tau_{dyn}$ does not vary as strongly with temperature 
as $\tau_{rad}$ and will therefore vary more mildly throughout HAT-P-2b's orbit.
We find that near the 100~mbar level of our simulations at periapse passage $\tau_{rad} <1$~hour and $\tau_{dyn}\sim10$~hours, which results in a 
large day-night temperature contrast during this phase of HAT-P-2bÕs orbit.  As HAT-P-2b approaches apoapse, $\tau_{rad}$ grows to be 
on the order of 5-6 hours while $\tau_{dyn}$ changes very little, which results in some muting of the day-night temperature difference.
This qualitatively explains the variations in the day/night temperature difference throughout the orbit of HAT-P-2b seen in 
Figures~\ref{hat2_temp_100mbar_noTiO} and  \ref{hat2_temp_100mbar_eqchem}.

An important regime shift in the global jet structure occurs as HAT-P-2b goes from the periapse to the apoapse of its orbit.  
At periapse, HAT-P-2b receives an amount of stellar insolation equivalent to that received by the `very-hot' Jupiter HAT-P-7b 
in its circular orbit.  At these high effective temperatures ($T_{eff}\sim2400$~K) the formation of an equatorial 
jet is suppressed and instead air heated near the substellar point flows uniformly outward toward the day/night terminator, forming a `day-to-night' circulation pattern 
(see bottom panels in Figures~\ref{hat2_temp_100mbar_noTiO} and \ref{hat2_temp_100mbar_eqchem}).  
The jet suppression is due to the fact that the radiative timescales at these high temperatures are much shorter than the 
time required for the Kelvin and Rossby waves that drive the super-rotating (eastward) equatorial jet to propagate a planetary radius \citep{sho11, sho13}.  

Near apoapse, the amount of incident stellar flux received by HAT-P-2b is similar to that received by the well-studied non-eccentric 
transiting hot-Jupiter HD~189733b.  A number of both theoretical \citep[e.g.][]{sho09, rau13} and observational \citep[e.g.][]{knu07, knu12} 
studies have sought to understand the atmospheric circulation regime of HD~189733b.  The consensus among both modelers and observers 
is that a super-rotating equatorial jet exists in HD~189733b's atmosphere, which causes an overall shift in the observed thermal pattern of the 
planet of $\sim$30$^{\circ}$ to the east of the substellar longitude.  The theory behind the formation of this super-rotating jet is 
well described in \citet{sho11}.  We see a similar eastward jet form near the equator in our HAT-P-2b simulations away from periapse 
(see top panels in Figures~\ref{hat2_temp_100mbar_noTiO}~and~\ref{hat2_temp_100mbar_eqchem}).  Our simulations predict that the variations in the incident flux on the planet due 
to its eccentric orbit causes it to shift between the jet dominated and day-night flow dominated regimes identified in \citet{sho13}.

\subsubsection{Equatorial Hot Spot}

In our simulation we see an interesting interplay between radiative and advective timescales that causes the displacement 
of the hottest point on the planet, or hotspot, from the substellar point to evolve with orbital phase.  Figure \ref{hat2_hot_spot}
shows the temperature, averaged in latitude, as a function of pressure and longitude from the 
substellar point near periapse and secondary eclipse for our 1$\times$ solar models both with and without TiO/VO.  
A clear difference between those models that have TiO/VO and those that do not is the pressures at which the 
hot spot predominately resides.  In our case without TiO/VO the hot spot is generally centered around the 1~bar 
level while our models with TiO/VO have a hot spot centered between 1 and 10~mbar.  This decrease in the 
pressure level of the hot spot is due to the fact that TiO and VO absorb incoming visible wavelength 
starlight at lower pressures, thus creating a thermal inversion and the vertical shift in the hot spot seen in 
Figure~\ref{hat2_hot_spot}.

Above the 10~bar level in our models near periapse (Figure~\ref{hat2_hot_spot}), we see that the hot spot in the models without TiO/VO is offset from the substellar point by 
about 12$^{\circ}$ while the hot spot is offset from the substellar point by less than 5$^{\circ}$ in our models with TiO/VO.  Although 
wind speeds are much greater in our models with TiO/VO (Figure~\ref{hat2_vrms_vs_time}), radiative timescales are much shorter compared 
to models that do not include TiO/VO due to the pressure levels at which the hot spot develops.  This means that parcels of air cool more rapidly than they can be advected downwind in our cases with TiO/VO, 
which result in a smaller offset in the hot spot compared to models that do not include TiO/VO.

\begin{figure*}
\centering
 \includegraphics[width=0.45\textwidth]{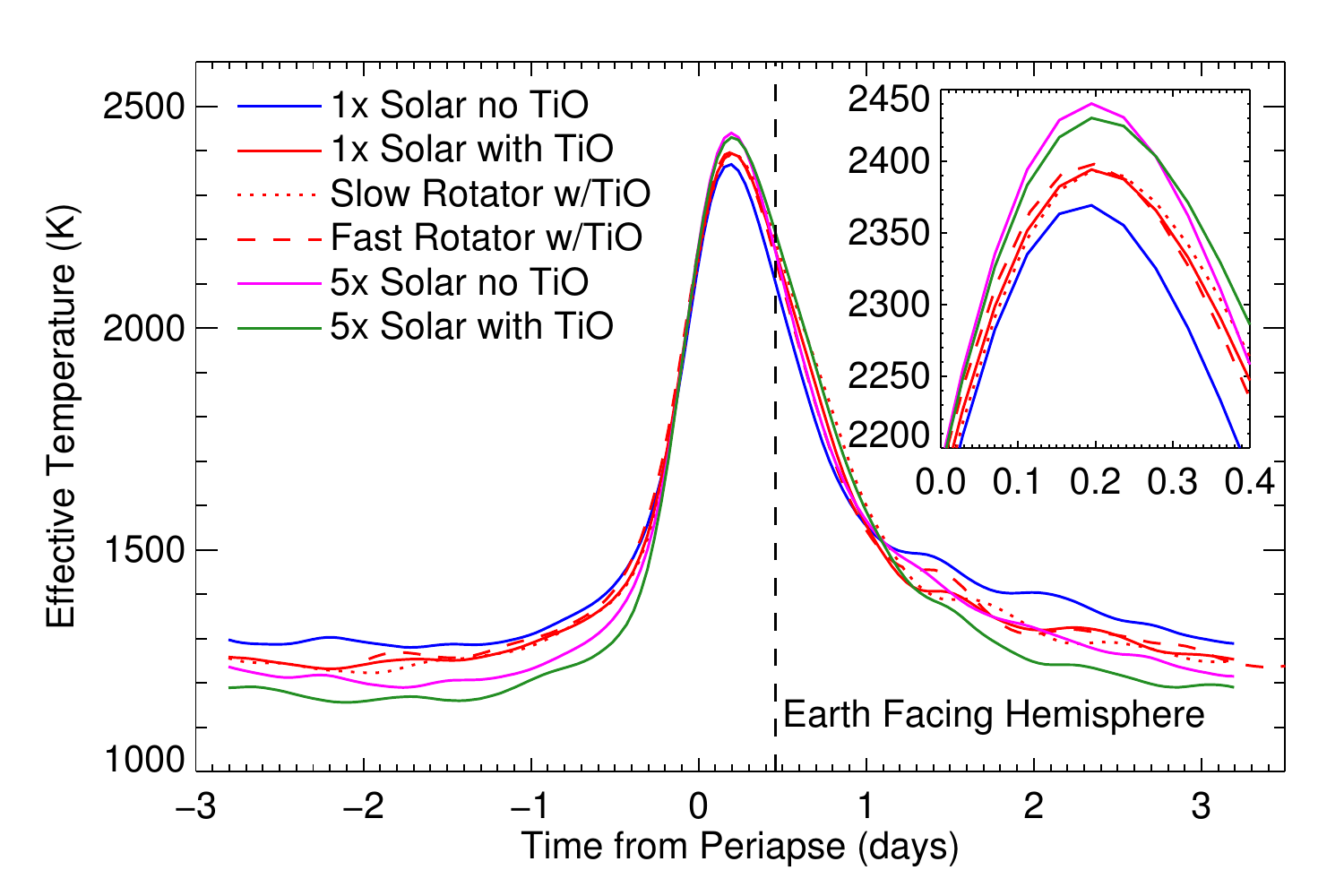}
 \includegraphics[width=0.45\textwidth]{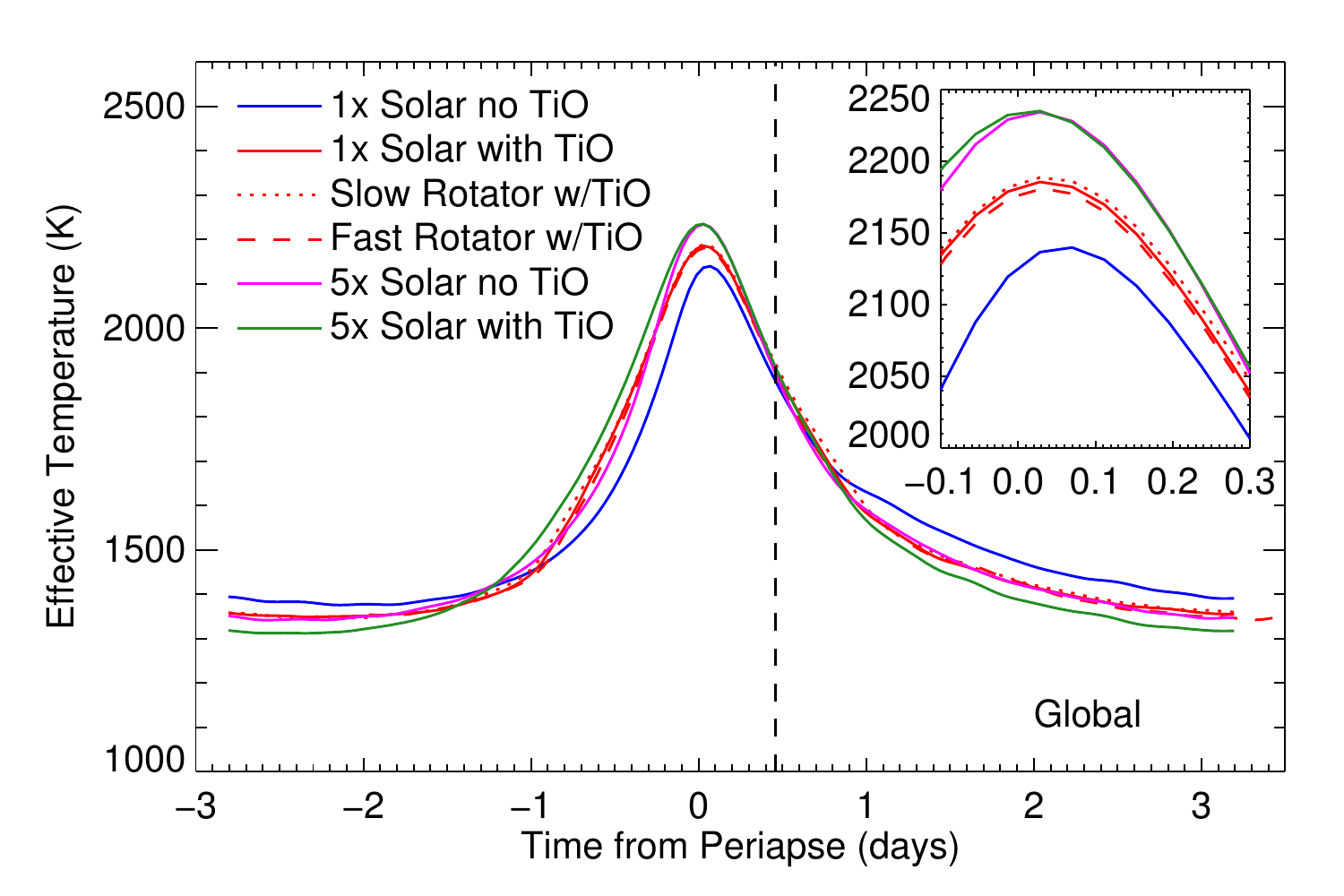}\\
 \includegraphics[width=0.45\textwidth]{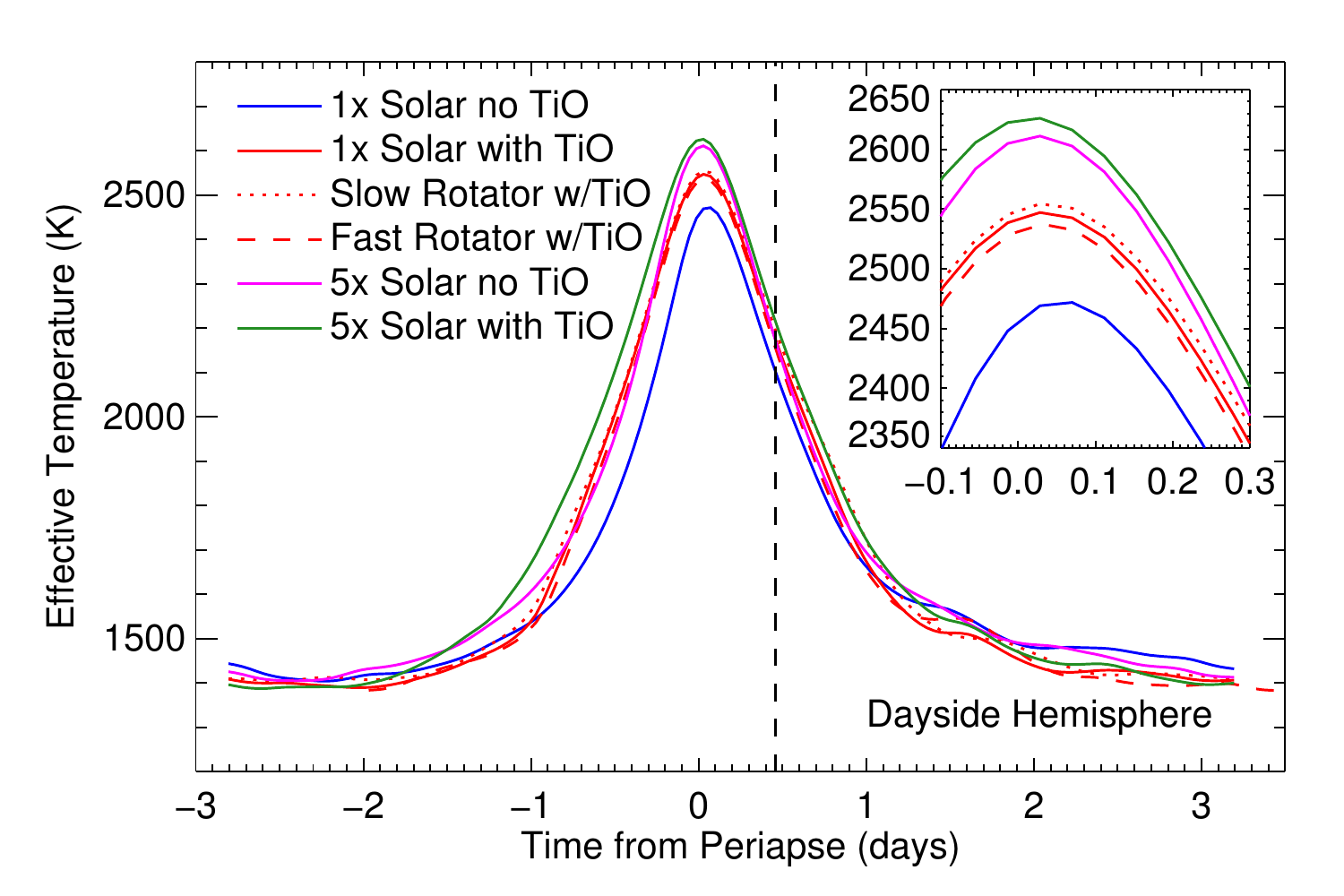}
 \includegraphics[width=0.45\textwidth]{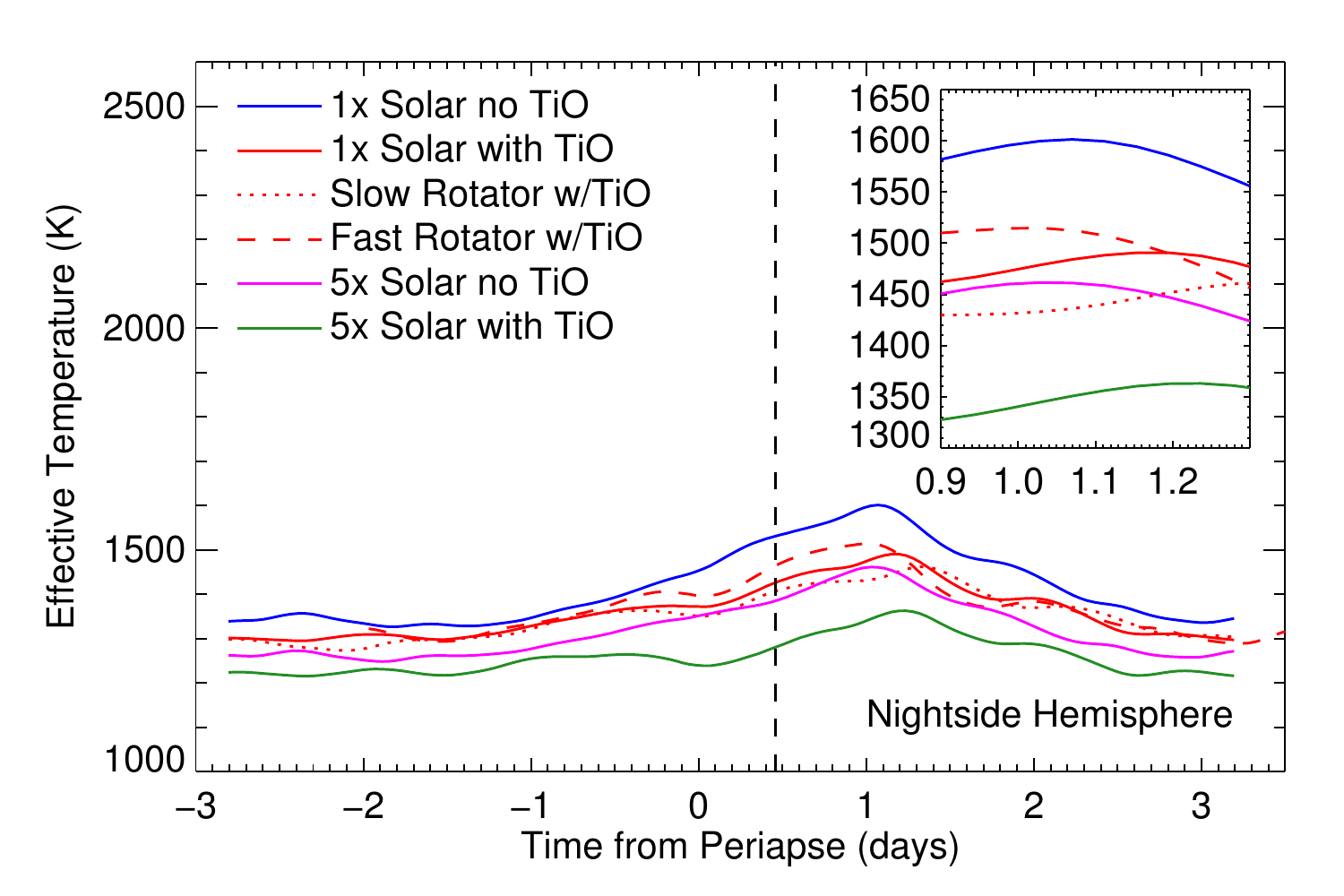}
  \caption{Effective temperature as a function of time from periapse passage for the earth-facing hemisphere (top left), entire planet (top right), 
                 dayside hemisphere (bottom left), and nightside hemisphere (bottom right) for 
                 each of the simulations considered here.  Insets provides a detailed view of the peak in effective temperature for each case, 
                 which highlights the difference in the magnitude and offset of the peak in planetary flux between each 
                 of the simulations.  Dashed vertical line in each panel represents the temporal location of secondary eclipse.
                  }\label{hat2_model_teff}
\end{figure*}

We also observe that the magnitude of the hotspot offset increases dramatically in the pressure region between 1 and 10 bar as the planet goes from 
perhaps to secondary eclipse.  This increase in the hotspot offset as a function of orbital phase occurs in both simulations with and without TiO/VO.   
It is in the 1 to 10 bar region of the atmosphere that radiative and advective timescales 
become similar and the efficient transport of `heat' deposited near periapase away from the substellar point can occur.  Wavelengths 
that probe this deep pressure region may show a pronounced offset in the peak of the thermal phase curve away from secondary eclipse and towards periapse 
given the geometry of the HAT-P-2 system.  This shift in the hot spot highlights the importance 
of self-consistent treatment of three-dimensional radiative and advective processes in atmospheric models for exoplanets, especially those on eccentric orbits.

\subsection{Theoretical Light Curves and Spectra} 

The SPARC model is uniquely equipped to produce both theoretical light curves and spectra directly from our three-dimensional 
atmospheric model for HAT-P-2b.  Once our simulations for HAT-P-2b reach an equilibrium state, we record pressure 
and temperature profiles along each grid column at many points along the planet's orbit (Figure~\ref{hat2_orb_fig}).  
These pressure-temperature profiles are then used in high resolution spectral calculations to determine the emergent flux from 
each point on the planet and which portion of that emergent flux would be directed toward an earth observer including
limb darkening/brightening effects.  Spectra and light curve generation methods are fully described in \citet{for06}.

The three-dimensional nature of our models also allows us to invoke any number of viewing angles and wavelength 
ranges in determining the theoretical emission from as a function of time.  In Figure~\ref{hat2_model_teff} 
we present the effective temperature of the earth-facing hemisphere, dayside, and global average as a function 
of time from periapse passage.  The effective temperature of a hemisphere is calculated by integrating the 
planetÕs emitted spectrum over all wavelengths from a particular viewing location such as Earth or the parent star, 
to obtain the total flux from the hemisphere. This quantity is then divided by the StefanÐBoltzmann constant, $\sigma$,  
and the fourth root it taken to obtain a temperature. For the global effective temperature calculation, the flux from both 
hemispheres (e.g., for day and night, far above the substellar point and anti-stellar point, respectively) are first added 
together before a temperature is determined.

The effective temperatures presented in Figure~\ref{hat2_model_teff} provide important insights into 
the planetary response to the variations in heating experienced by HAT-P-2b throughout its orbit from our various models.
The effective temperatures for the earth-facing hemisphere clearly show a trend between the magnitude of the peak 
effective temperature and the time lag between periapse passage and when the peak in temperature occurs.
This trend is predominately the result of the orbital geometry of the 
HAT-P-2 system as the opposite trend in peak timing and magnitude can be seen in the dayside effective 
temperatures in the bottom left panel of Figure~\ref{hat2_model_teff}.  The trend in the dayside hemisphere 
effective temperatures results from opacity differences between the models that exist because of changes in the dayside 
thermal structure and composition based on our equilibrium chemistry assumptions.  As can been seen in the 
evolution of the nightside effective temperature as a function of time from periapse passage, those models that 
exhibit the greatest dayside effective temperatures also have the lowest nightside effective temperatures, which is 
the result of `inefficient' day-to-night heat transport at photospheric pressures.  Interestingly, the only clear 
deviations between the effective temperatures of the nominal 1$\times$ solar with TiO case and its fast and 
slow rotating counterparts are seen 
in the nightside effective temperature as a function of orbital phase.  It is important to keep in mind that 
the assumed nominal rotation period of HAT-P-2b is $\sim$1.95~days and that the timing of the peak in the 
nightside effective temperature is not just the result of the region heated at periapse rotating onto the 
nightside hemisphere.  Instead the timing in the peak nightside temperature is the result of the evolution of 
the planet's thermal and wind structure as discussed in Section~\ref{evolution}.

\begin{figure}
\centering
 \includegraphics[width=0.5\textwidth]{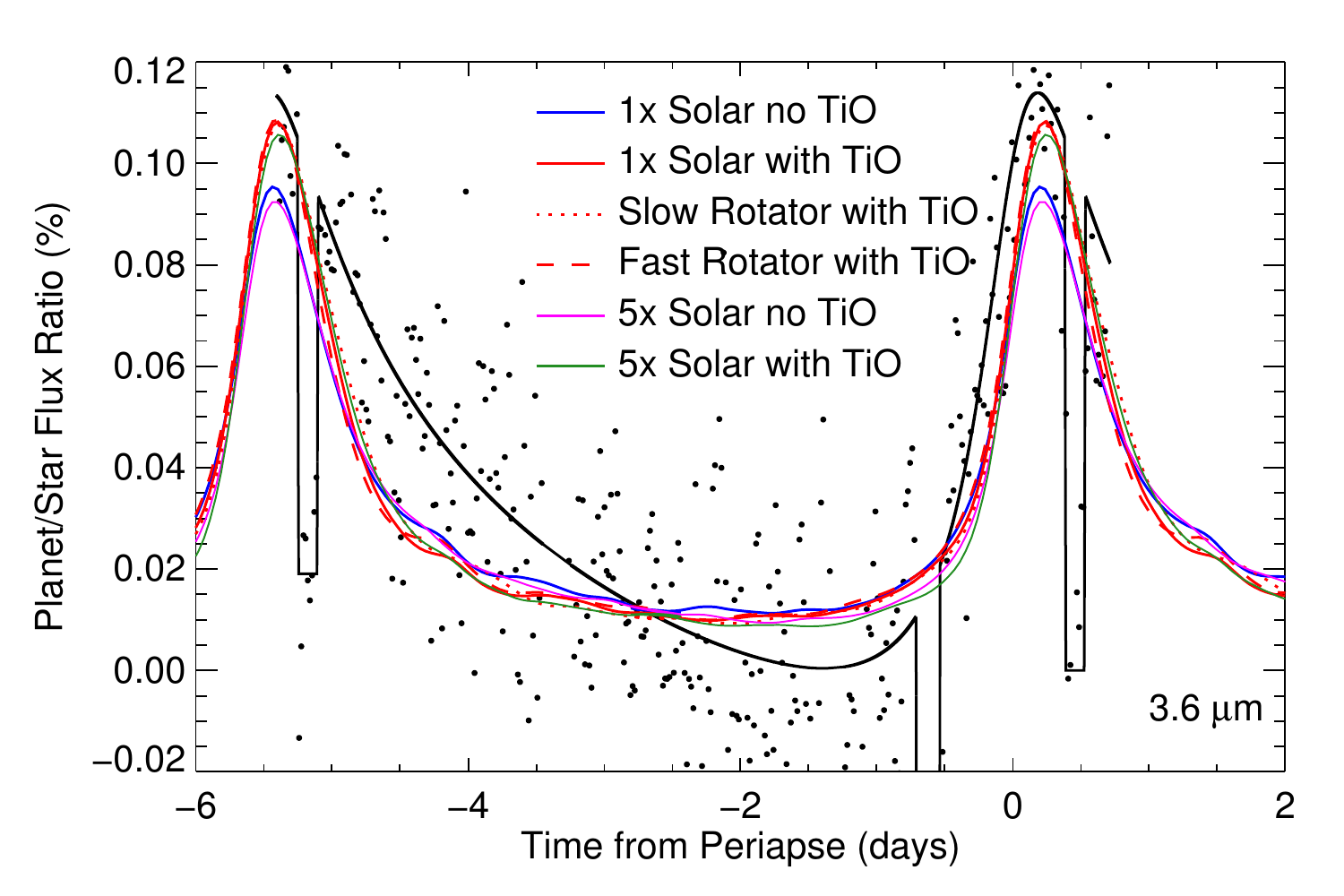}\\
 \includegraphics[width=0.5\textwidth]{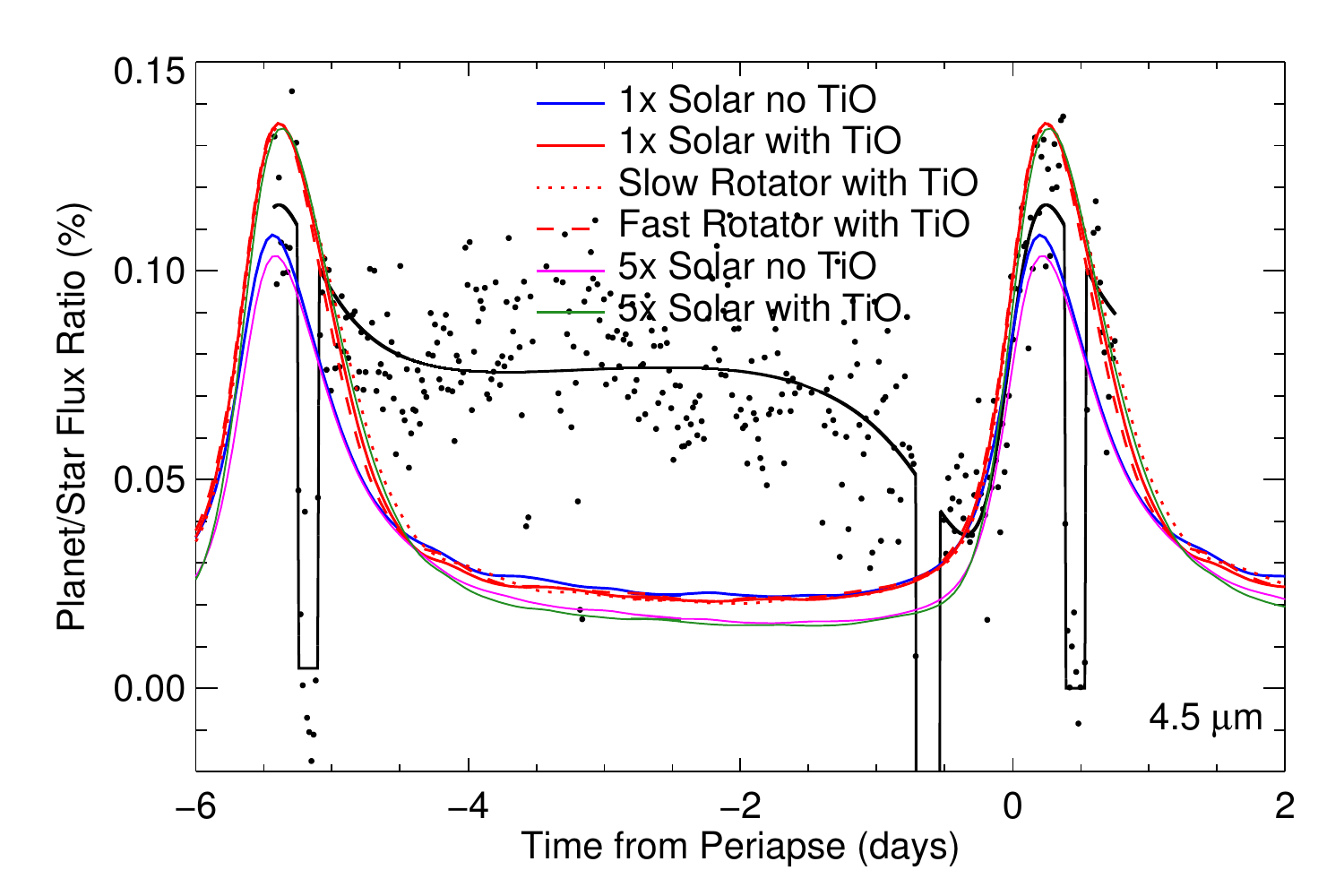}\\
 \includegraphics[width=0.5\textwidth]{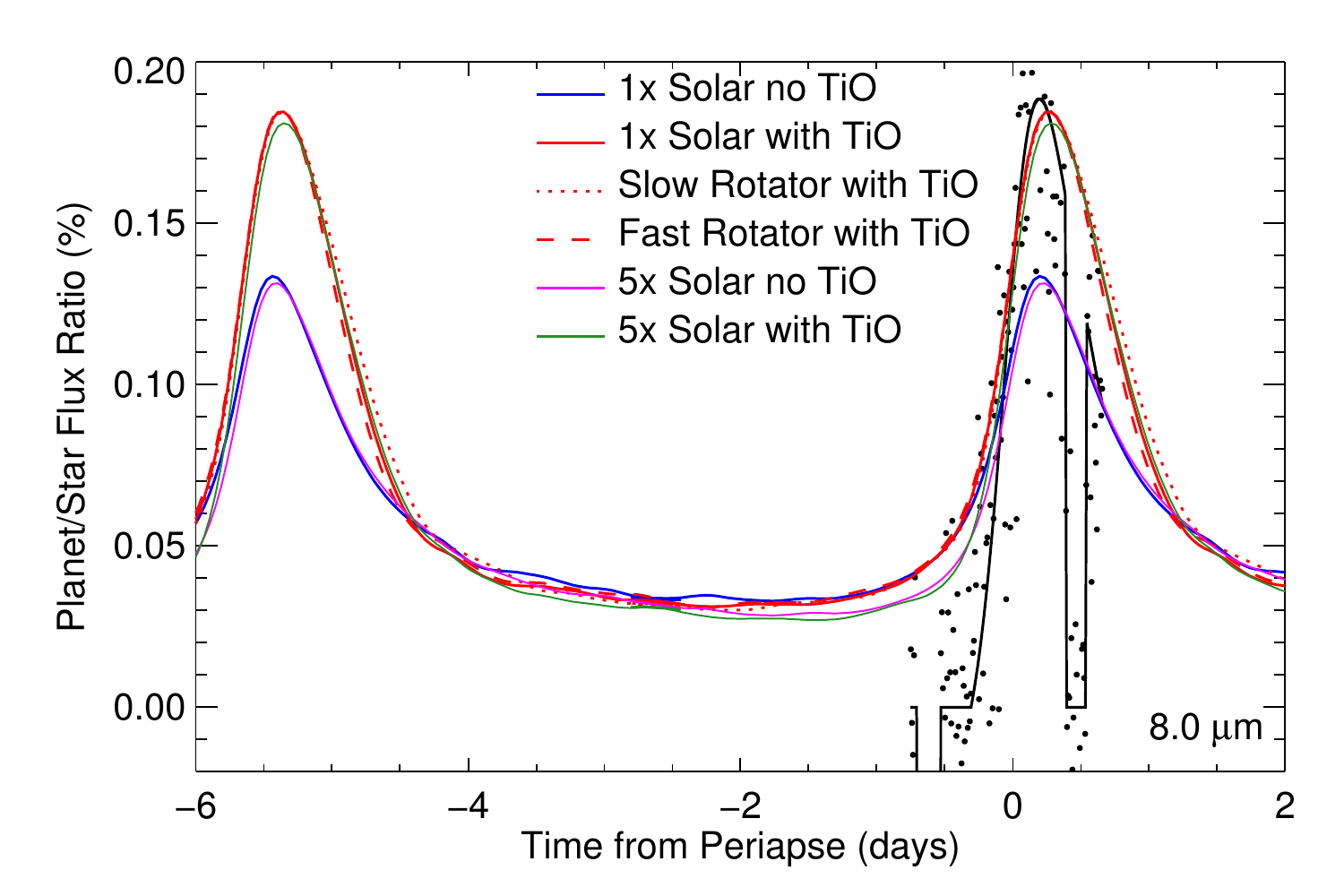}
  \caption{Theoretical planet/star flux ratio as a function of time from periapse for the 3.6 (top), 4.5 (middle), 
                  and 8.0~$\mu$m (bottom) {\it Spitzer} bandpasses compared with the observed light curves from \citet{lew13}.
                  The observations have been binned into 20 minute intervals.  For clarity, the solid black lines show 
                  the best-fit phase curves from \citet{lew13}.
                  }\label{hat2_model_lc}
\end{figure}

Figure \ref{hat2_model_lc} shows our theoretical light curves, expressed as a planet/star flux ratio 
as a function of time from periapse passage, compared with the observed {\it Spitzer} light curves 
at 3.6, 4.5, and 8.0~$\mu$m.  The strong peaks that we see in our theoretical light curves is the result of a 
combination of intense heating at periapse and the offset (or lack thereof) of the hotspot from the substellar longitude.    
If HAT-P-2b's thermal structure did not vary with orbital phase and there were no shift in the hotspot from the substellar longitude, 
then we would expect the peak of the light curve to occur near secondary eclipse. 
As shown in Figure~\ref{hat2_hot_spot}, the hot spot in HAT-P-2b's atmosphere is shifted eastward from the substellar longitude in both the cases with 
and without TiO/VO.  This shift in the hot spot means that 
the hottest portion of the planet will rotate into view a few hours ($\sim 2$~hours) before secondary eclipse
($\sim8-11$~hours after periapse passage).   The timing of the peaks of our theoretical light curves represents an average value 
that lies between the peak determined by radiative timescales and the peak determined by dynamical timescales (orbital geometry, rotation, and winds).   

It is clear that significant differences exist between the various model light curve predictions 
and between the observations and the model light curve predictions as a whole.  The light curve predictions from 
models that include TiO/VO are fairly insensitive to variations in both composition (1$\times$ vs. 5$\times$ solar 
metallicity) and rotation rate (0.5$\times$ and 2$\times$ nominal rotation period).  Models that do not include TiO/VO, 
and therefore do not have a strong dayside inversion near periapse, give theoretical light curves that underestimate the 
planetary flux during the majority of planetary orbit at 3.6, 4.5, and 8.0~$\mu$m, especially in the region near 
periapse.  It is important to note that if phase curve observations of HAT-P-2b were made in only one bandpass, that 
entirely different conclusions about the properties of the planet's atmosphere would be drawn.  For example, if the 
3.6 or 8~$\mu$m observations were taken on their own, or even combined, a comparison with the theoretical 
phase curves would strongly favor a scenario with a solar, or slightly super-solar, composition atmosphere in chemical 
equilibrium with a dayside thermal inversion, possibly 
caused by the presence of TiO/VO.  However, the 4.5~$\mu$m observations do not conform particularly well to 
any of the theoretical light curves from our various models.  This points to some physical process that is currently 
missing from our models and highlights the importance of multi-wavelength observations when constraining the 
properties of exoplanet atmospheres.

\begin{figure}
\centering
 \includegraphics[width=0.5\textwidth]{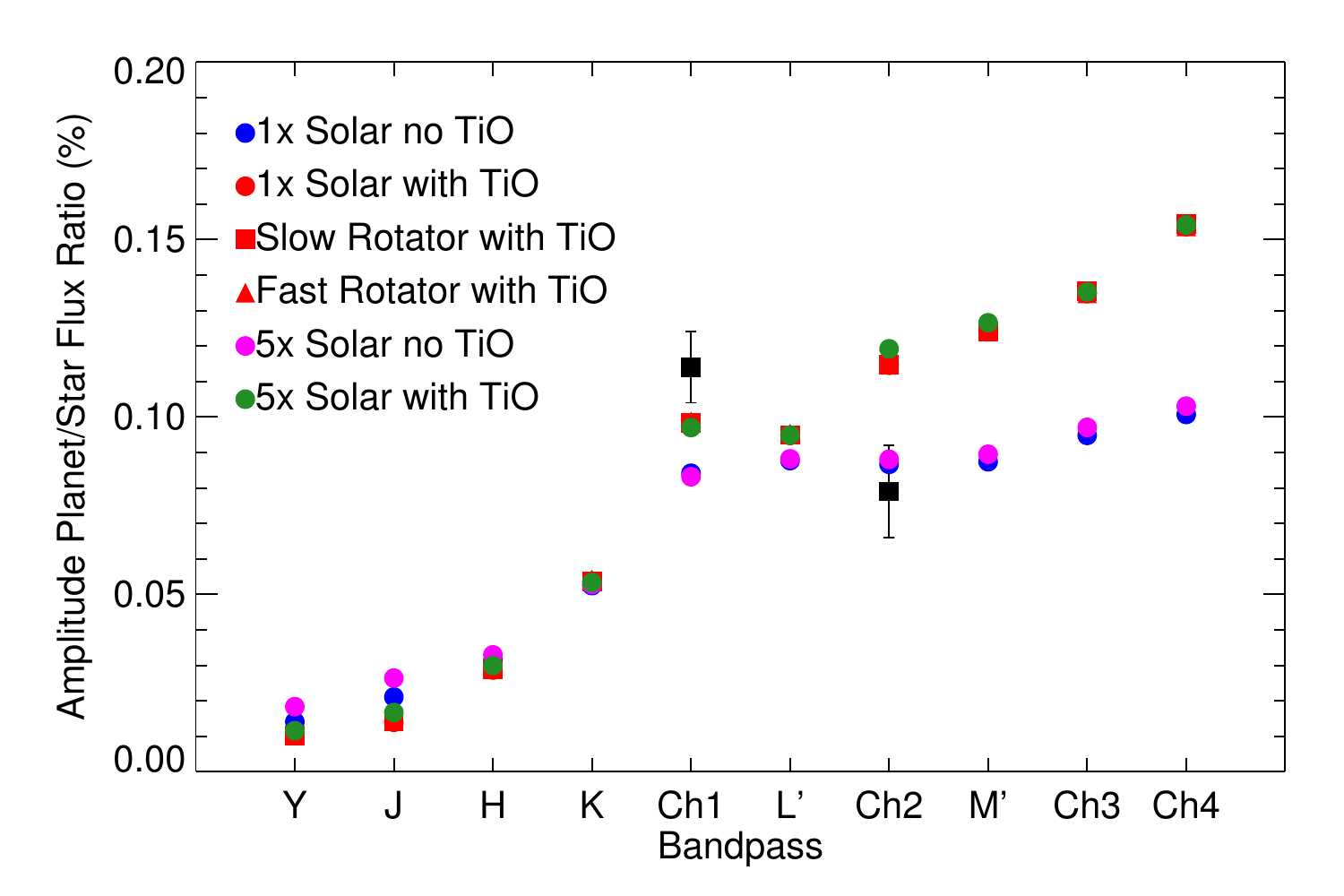}\\
 \includegraphics[width=0.5\textwidth]{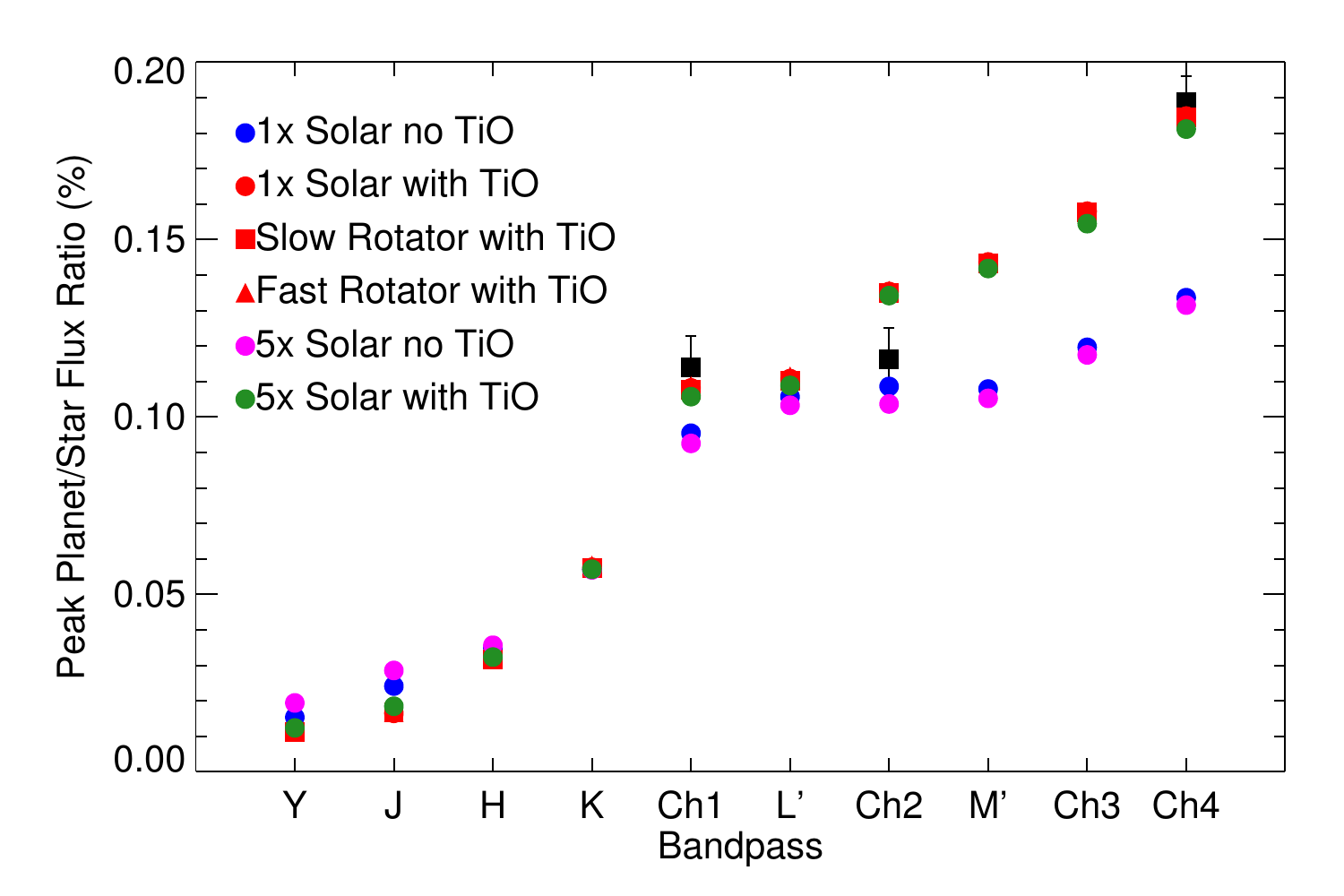}\\
 \includegraphics[width=0.5\textwidth]{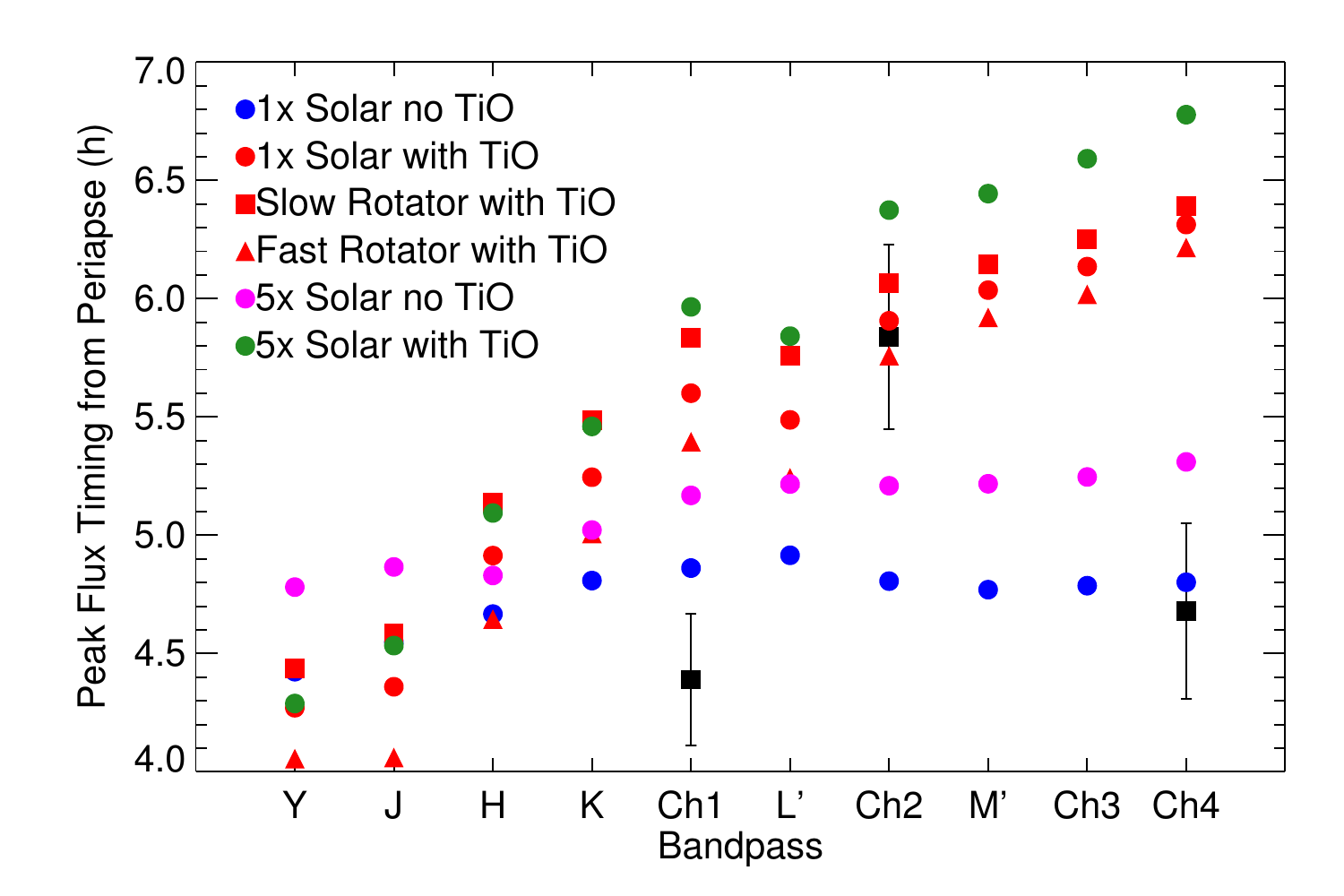}
  \caption{Amplitude (top), peak (middle), and timing of the peak (bottom) of the planet/star flux ratio as a function of bandpass.  Ch1, Ch2, Ch3, and Ch4 represent 
  the 3.6, 4.5, 5.8, and 8.0~$\mu$m bandpasses of the {\it Spitzer} IRAC instrument respectively.  The ground-based Y, J, H, K, L$\prime$, and M$\prime$ bandpasses 
  have central wavelengths of 1.02, 1.26, 1.62, 2.21, 3.78, and 4.78~$\mu$m respectively.
  Observed values with one sigma error bars from \citet{lew13} are 
  over plotted with the black squares.}\label{hat2_model_comp}
\end{figure}

The amplitude (difference between maximum and minimum of planetary flux) of the phase curves along with the maximum planetary flux value and the timing 
of that maximum from periapse reveal a 
great deal about atmospheric radiative and dynamical timescales in an eccentric planet's atmosphere.  Figure~\ref{hat2_model_comp} 
shows how these key phase curve parameters vary as a function of bandpass and model parameters and compares them with the values 
derived for the 3.6, 4.5, and 8.0~$\mu$m {\it Spitzer} light curves from \citet{lew13}.  There is a clear divergence between models that incorporate TiO/VO 
into the opacity tables and those that do not, especially at longer wavelengths.  It is also interesting that varying the rotation rate of the planet by a factor 
of two only results in small changes in the timing and shape of the peak in the planetary flux.  Given the one sigma error bars from the {\it Spitzer} observations, it is 
impossible to distinguish the slow and fast rotation period models from the nominal pseudo-synchronous rotation period model.  This highlights that 
atmospheric radiative timescales and the viewing geometry of the system predominately determine the shape of HAT-P-2b's phase curve.  

\begin{figure}
\centering
 \includegraphics[width=0.5\textwidth]{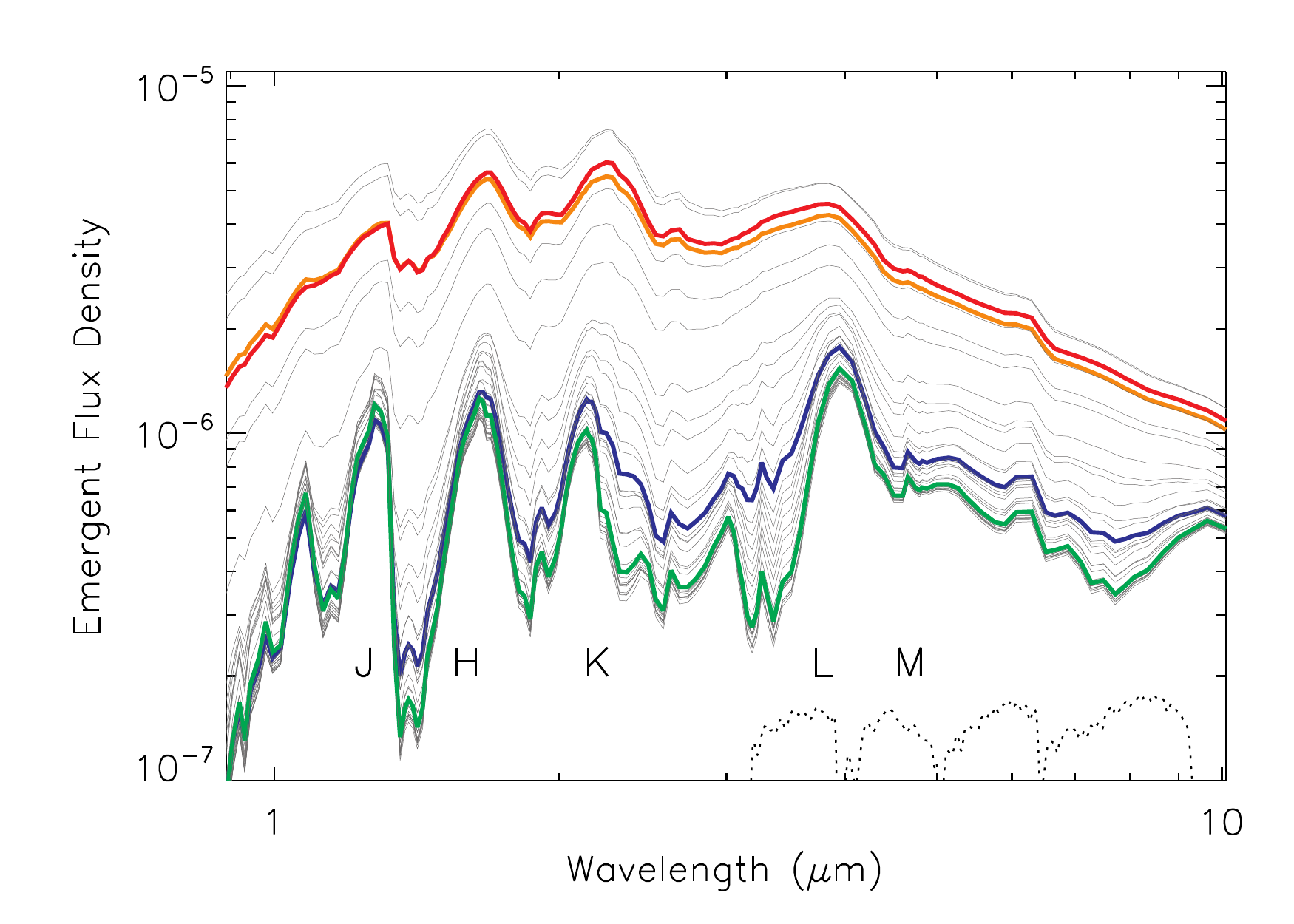}\\
 \includegraphics[width=0.5\textwidth]{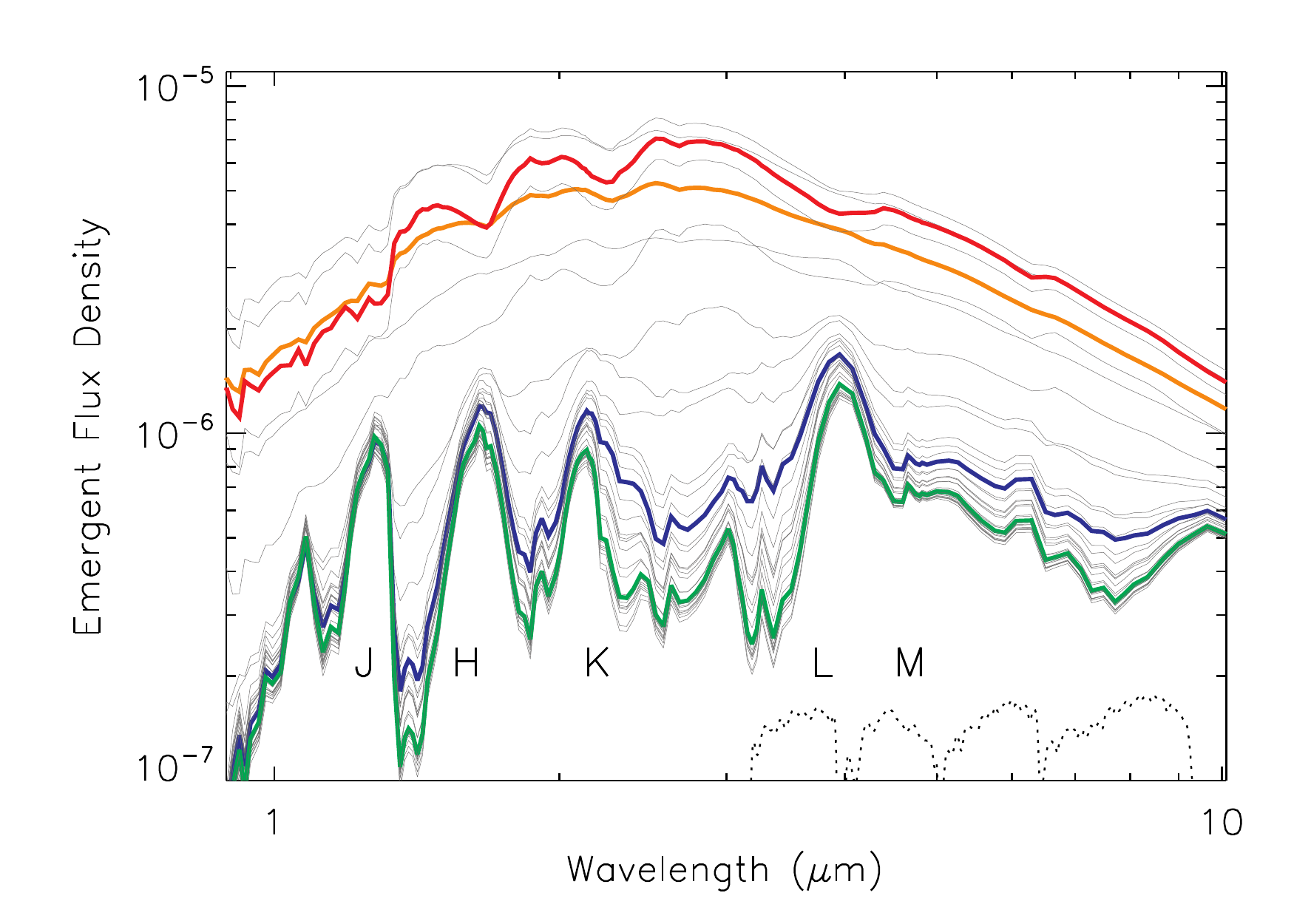}
  \caption{Flux per unit frequency, $F_{\nu}$ (erg s$^{-1}$ cm$^{-2}$ Hz$^{-1}$), as a function of wavelength for our solar metallicity 
  	atmospheric model of HAT-P-2b with (bottom) and without (top) TiO/VO.  The spectra are taken from a subset of points along a single orbit as indicated in Figure 
	\ref{hat2_orb_fig}.  The central wavelengths of J-, H-, K-, L-, and M-bandpasses are indicated by their corresponding letters. 
	Dotted lines at the bottom indicate the bandpasses of the four {\it Spitzer} IRAC bands from 3-9 $\mu$m. Colored spectra 
	are taken near secondary eclipse (red), periapse (orange), transit (blue), and apoapse (green).  Note how the absorption features in the spectra from the 
	 model with TiO/VO change to emission features as the planet approaches periapse.}\label{hat2_model_flux}
\end{figure}

If we plot the thermally emitted flux from the Earth-facing hemisphere of HAT-P-2b as a function of wavelength (Figure \ref{hat2_model_flux}) 
a number of features are readily apparent.  Most 
notably, the fluxes at secondary eclipse (red) and periapse (orange) are similar.  While the planet is hotter at periapse than at secondary 
eclipse, at periapse we do not see the fully illuminated hemisphere. The black lines are spaced evenly in time throughout the orbit
demonstrating that the planet spends most of its time at larger orbital separations where the thermal emission is weaker.

\begin{figure*}
\centering
\includegraphics[width=0.24\textwidth]{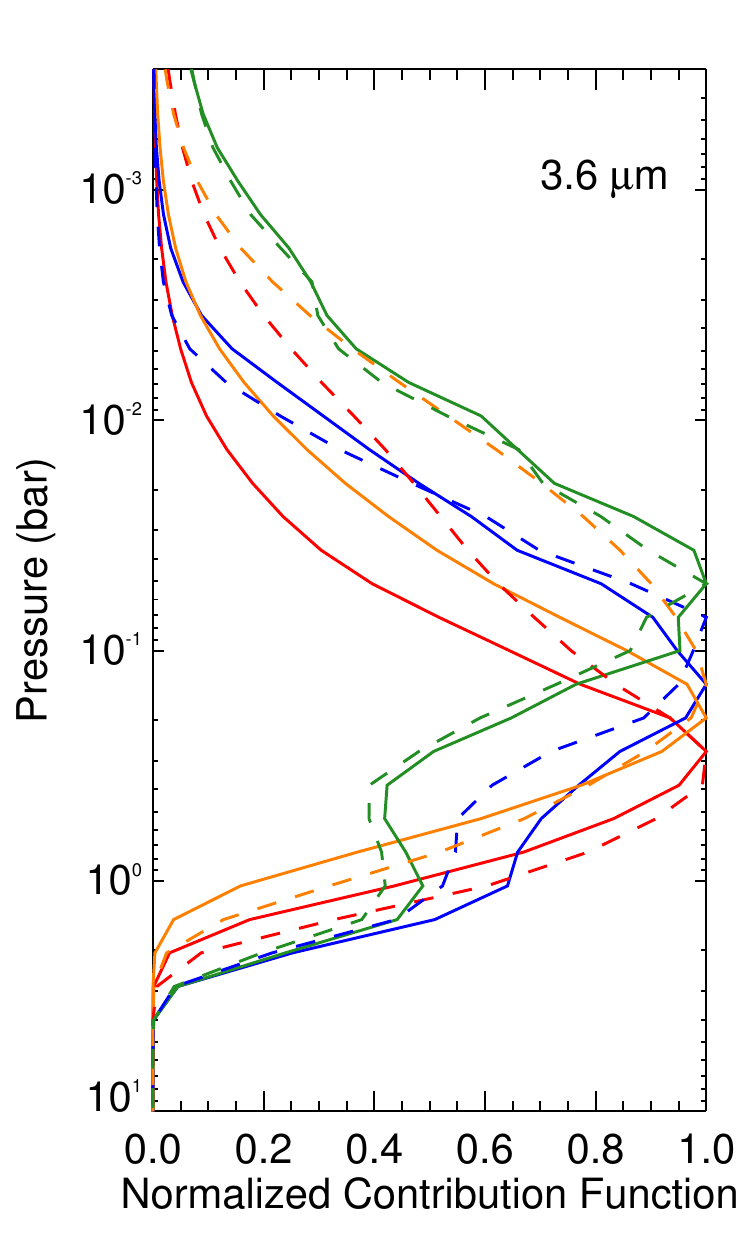}
\includegraphics[width=0.24\textwidth]{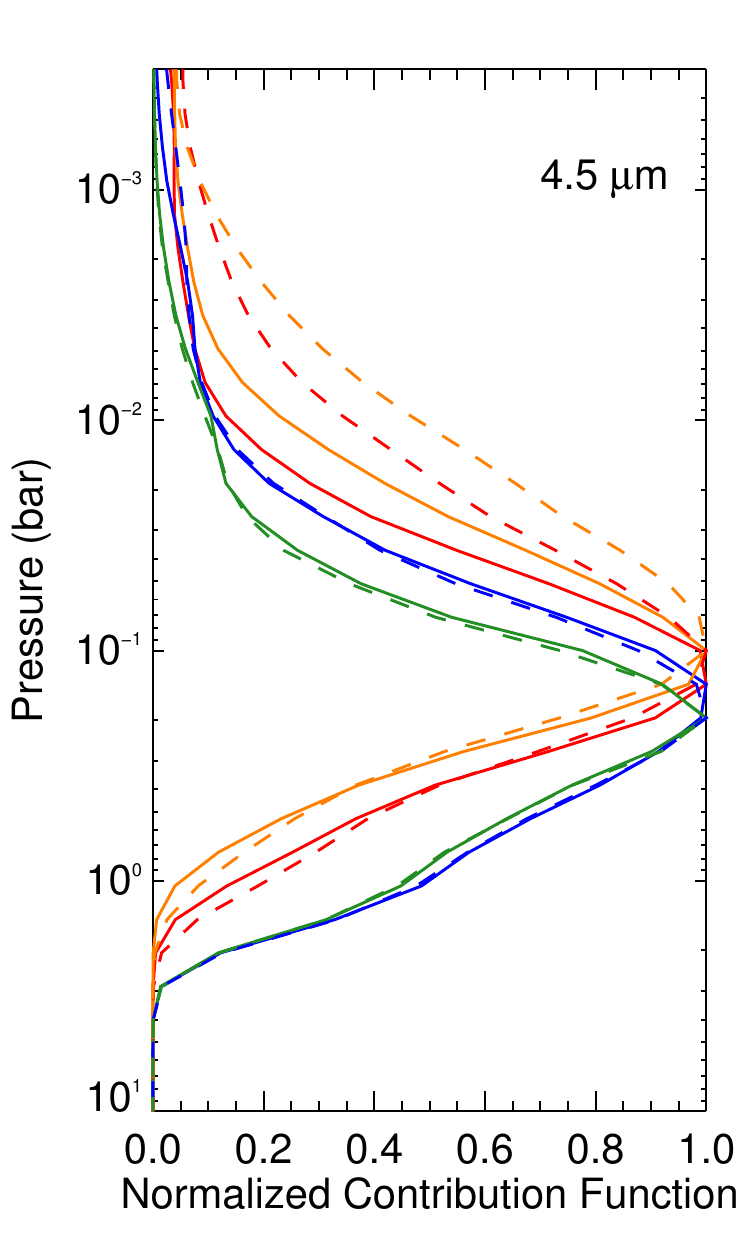}
\includegraphics[width=0.24\textwidth]{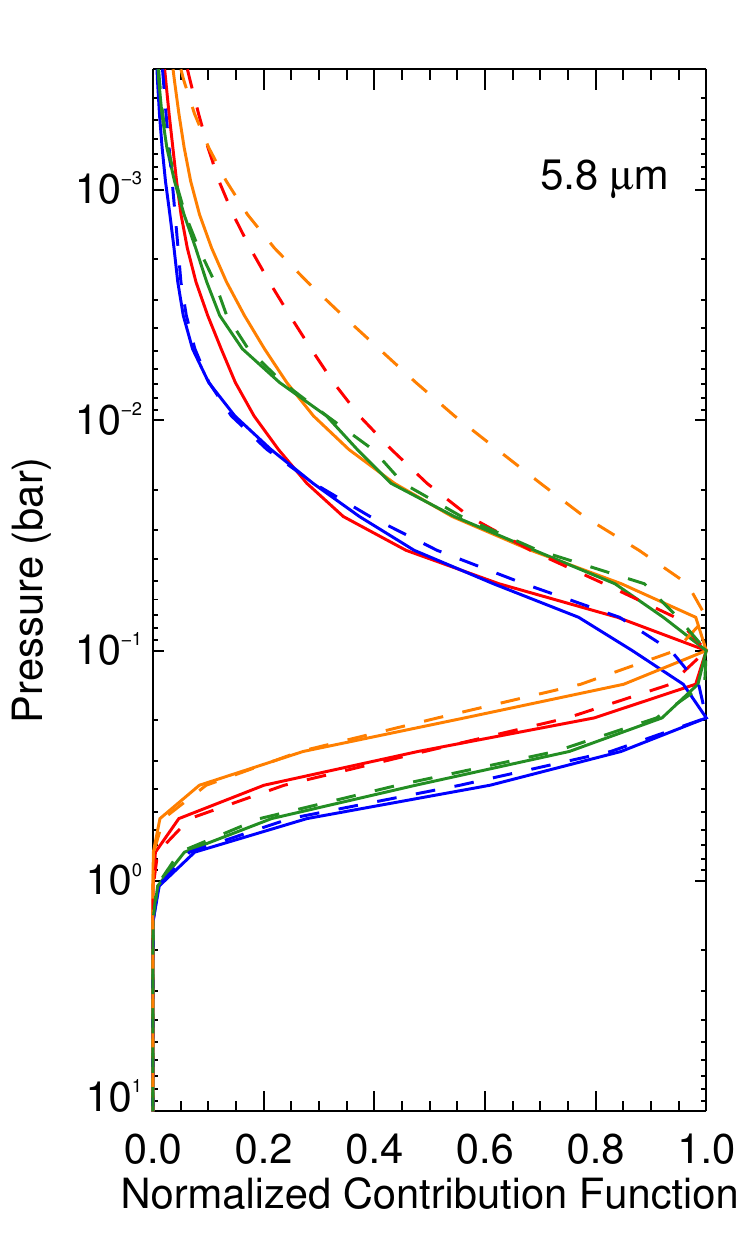}
\includegraphics[width=0.24\textwidth]{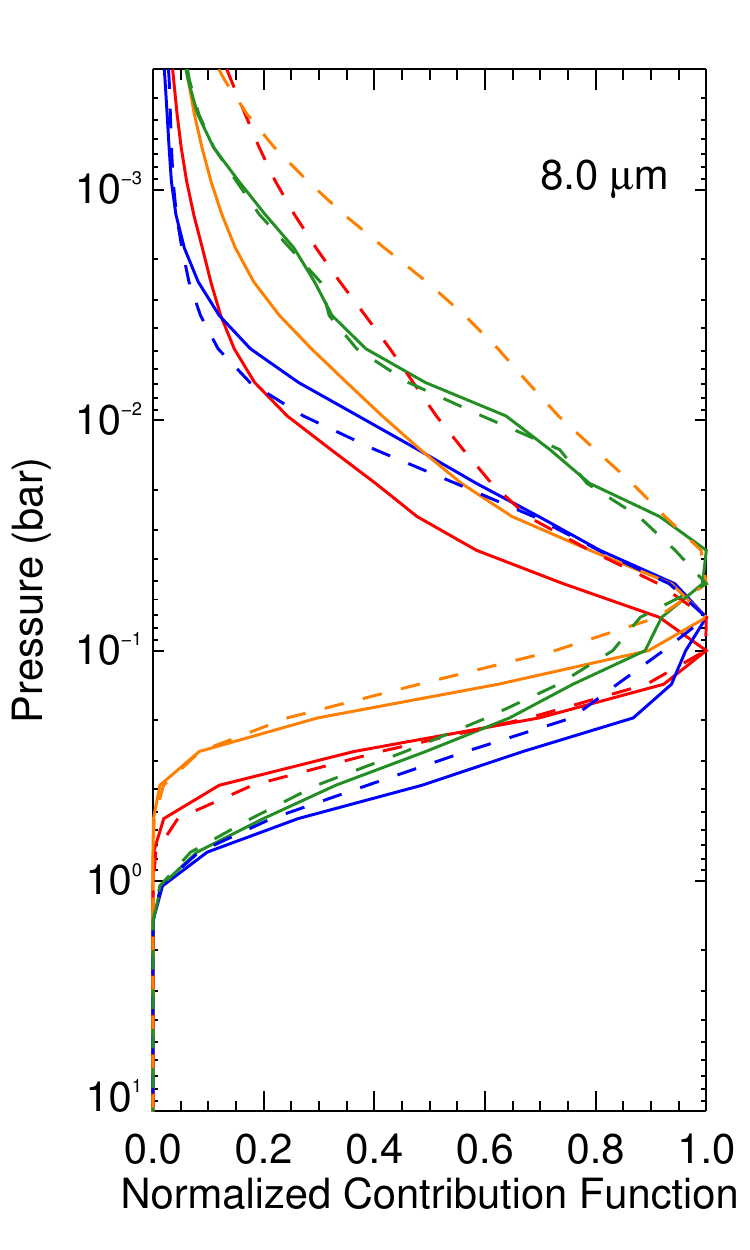}
\caption{The normalized contribution of each pressure level in our 1$\times$ solar models to the 3.6, 4.5, 5.8, and 8.0~$\mu$m planetary flux at periapse (orange), 
                secondary eclipse (red), apoapse (green), and transit (blue).   Solid lines are for our model that did not incorporate TiO/VO into the opacity tables, while 
                the dashed lines are for our model that did include TiO/VO. }\label{hat2_cf_plots}
\end{figure*}

One of the key advantages of the SPARC model is that we can consider how the contribution of flux from a given pressure level in our model 
might evolve with the planet's orbit \citep[e.g.][]{knu09a}.  This is especially important for planets on eccentric orbits where large changes in incident flux and
atmospheric chemistry of the planet are expected.  Figure~\ref{hat2_cf_plots} shows the normalized contribution function of the emitted planetary flux  
at transit, periapse, secondary eclipse, and apoapse for each of the {\it Spitzer} IRAC bandpasses.  These contribution functions are computed from 
a single pressure temperature profile that represents an average over the visible hemisphere of the planet that is weighted by the viewing angle 
\citep[see][for a discussion of the viewing angle]{for06}.

It is important to note how the pressure levels probed by a given bandpass
evolves with orbital phase and the difference in the contribution functions between models that include TiO/VO and those that do not (Figure~\ref{hat2_cf_plots}).  
The evolution of the 3.6~$\mu$m contribution functions compared with the 4.5~$\mu$m contribution highlights changes in the relative abundances of CO and 
CH$_4$ in our models of HAT-P-2b's atmosphere as a function orbital phase and global temperatures.  
The 4.5~$\mu$m bandpass probes the absorption features from 
CO in the $\sim$4.5-5.0~$\mu$m range while the 3.6~$\mu$m bandpass probes the strong absorption feature from CH$_4$ at 3.3~$\mu$m.
Differences between the pressures probed by models with and without TiO/VO in their atmospheres is, not surprisingly, strongest near periapse/secondary eclipse and 
enhances the contribution from lower pressure levels where the temperature inversion occurs and the planetary hot spot is well aligned with the substellar longitude 
(Figure~\ref{hat2_hot_spot}).  We expect the orbital variations in the contribution functions presented here will be reduced by chemical quenching \citep{coo06, vis12}, which 
is not currently accounted for in our models.  

\subsection{Variability} 

Typically when predicting the theoretical flux from the planet as a function of orbital phase from our three-dimensional simulations, 
we only consider a temporally averaged planetary thermal structure, or in the case of planets on eccentric orbits,  the thermal 
structure of the planet that develops during a single orbit.  As discussed in \citet{sho09} in the case of HD~189733b, orbit-to-orbit and 
longer term variations in an exoplanet's thermal structure are expected due to global scale oscillation and smaller scale turbulent processes.  
\citet{lan08} noted large variations in the thermal structure of their two-dimensional simulations of HAT-P-2b, but did not quantify how 
these thermal variations might manifest themselves observationally from one epoch to the next.  In Table~\ref{ecl_var}, we present the 
variations in the predicted eclipse depth as a function of bandpass from our nominal solar metallicity simulation that does not 
include TiO/VO.  These variations were calculated over a 100 simulated day period after the simulation had reached a stable configuration and 
include 24 estimates of the eclipse depth.  

\begin{deluxetable}{lcccc}
\tabletypesize{\scriptsize}
\tablecaption{HAT-P-2b Secondary Eclipse Depth Variability\label{ecl_var}}
\tablewidth{0pt}
\tablehead{
\colhead{Bandpass\tablenotemark{a}} & \colhead{$\lambda$\tablenotemark{b} ($\mu$m)} & 
\colhead{$\sigma$\tablenotemark{c}} & \colhead{$\Delta_{max}$\tablenotemark{d}} & 
\colhead{$\overline{P}_{band}$\tablenotemark{e} (bar)}
}
\startdata
Y &  1.02 & 5.4\% & 29.6\% & 2.13 \\   
J &  1.26 & 4.6\% & 25.0\% & 1.75\\ 
H &  1.62 & 2.1\% & 12.2\% & 0.78\\
K &   2.21 & 1.0\% & 4.5\% & 0.40\\
Ch1 & 3.6 &  0.8\% & 3.5\% & 0.33\\
L' &   3.78 &  0.9\% & 3.7\% & 0.12\\
Ch2 & 4.5 &  0.6\%  & 2.5\% & 0.18\\
M' &  4.78 &  0.6\% & 2.5\% & 0.05\\
Ch3 &  5.8 &  0.6\% & 2.7\% & 0.11\\
Ch4 &  8.0 &  0.6\% & 2.8\% & 0.09
\enddata
\tablenotetext{a}{Ch1, Ch2, Ch3, and Ch4 represent the four {\it Spitzer} IRAC bandpasses.  Y, J, H, K, L', and M' represent 
                            standard near- to mid-infrared bandpasses used at ground-based observatories.}
\tablenotetext{b}{Central wavelength of each bandpass.}
\tablenotetext{c}{Standard deviation of the theoretical eclipse depths normalized to the mean.}
\tablenotetext{d}{Maximum deviation in the theoretical eclipse depths normalized to the mean.}
\tablenotetext{e}{Average pressure probed by each bandpass as determined from the relevant contribution functions (See Figure~\ref{hat2_cf_plots}).}                      
\end{deluxetable}

A clear trend in the data presented in Table~\ref{ecl_var} is that the expected level of eclipse depth variability 
is strongly correlated with wavelength.  At the longer infrared wavelengths, the level of variability in the 
predicted eclipse depth is expected to be on the order of $\sim$1\%.  This is in line with the estimates for HD~189733b's
eclipse depth variability at 8~$\mu$m presented in \citet{sho09}.  At the shorter infrared wavelengths, the predicted 
variability in the eclipse depth grows reaching between 5\% and 10\%.  The level of variability in our simulations is also 
strongly correlated with the average pressure level probed by a given bandpass (last column Table~\ref{ecl_var}).  At depth in an 
exoplanet atmosphere,  the radiative and dynamical (including wave phenomenon) timescales grow longer and become commensurate
with the planetary orbital/rotational period near the 1 bar level of the atmosphere \citep{sho08}.  This means that variations in the thermal 
structure can develop on periods greater than the orbital period leading to a greater level of predicted variability in the secondary eclipse 
depth at shorter wavelengths.  We do not note any coherent oscillations in the predicted eclipse depth, but these oscillations may be at 
timescales less than an orbital period or greater than 24 orbital periods and would therefore not be readily apparent in our 
current analysis.  Although we only present estimates of eclipse variability from our solar metallicity simulation that 
does not include TiO/VO, we expect the same general trend of variability increasing with pressure probed by a given 
bandpass for all of our simulations and cloud-free exoplanet atmospheres in general.

\section{Discussion}\label{hat2_discussion}

In the previous section we have outlined the basic thermal and wind structures that develop in our atmospheric 
models for HAT-P-2b.  We have also compared theoretical light curves derived from our models to the observed 
light curves at 3.6, 4.5, and 8.0~$\mu$m from \citet{lew13}.  In \citet{lew13},  we investigated the range of 
radiative and advective timescales that would explain the magnitude and timing in the of the peak in the planetary 
flux using the semi-analytic model of \citet{cow11}.  The models presented here support our predictions from \citet{lew13} that 
radiative timescales are short ($\sim 2$~hours) and zonal wind speeds are large ($\sim 4-5$~km~s$^{-1}$) for 
HAT-P-2b near the periapse of its orbit as determined using the simplified models of \citet{cow11}.  
Our three-dimensional simulations allow us to more rigorously explore the interconnectivity between 
radiative and advective processes and provide additional context for the physical origin of phase curve variations we
observe and predict for HAT-P-2b.

We can compare the amplitude, magnitude, and timing of the peak in the planetary flux from our models in 
much the same way as was done in \citet{lew13} using the \citet{cow11} models.  From Figure~\ref{hat2_model_comp} 
it is clear that no one model provides the `best-fit' to the data and that at the three-sigma level most of our 
model predictions agree with the basic observed properties of the HAT-P-2b phase curve.  The observed peak flux at 8.0~$\mu$m (Ch4) 
is the only point that strongly favors models that have an atmospheric inversion due to the presence of TiO/VO versus those 
models that do not include TiO/VO.  
Interestingly, the 8.0~$\mu$m peak flux timing more strongly favors those models without 
an inversion, but it is possible that the exact timing of the peak of the flux may be more strongly influenced by the timescales for vertical and 
horizontal mixing of TiO/VO, which are not rigorously accounted for in this study.   
Although our models seem to capture the correct heating timescales observed at 3.6 and 4.5~$\mu$m, they underestimate the 
cooling timescale of the planet at 3.6 and 4.5~$\mu$m (Figure~\ref{hat2_model_lc}).  At 8~$\mu$m our models overestimate both the heating 
and cooling timescales of the planet.  These discrepancies between the observed and modeled heating/cooling timescales
point to changes in the pressure levels probed by each bandpass that are likely due to opacity (chemical) changes 
not captured by our model.  

\begin{figure*}
\centering
 \includegraphics[width=0.45\textwidth]{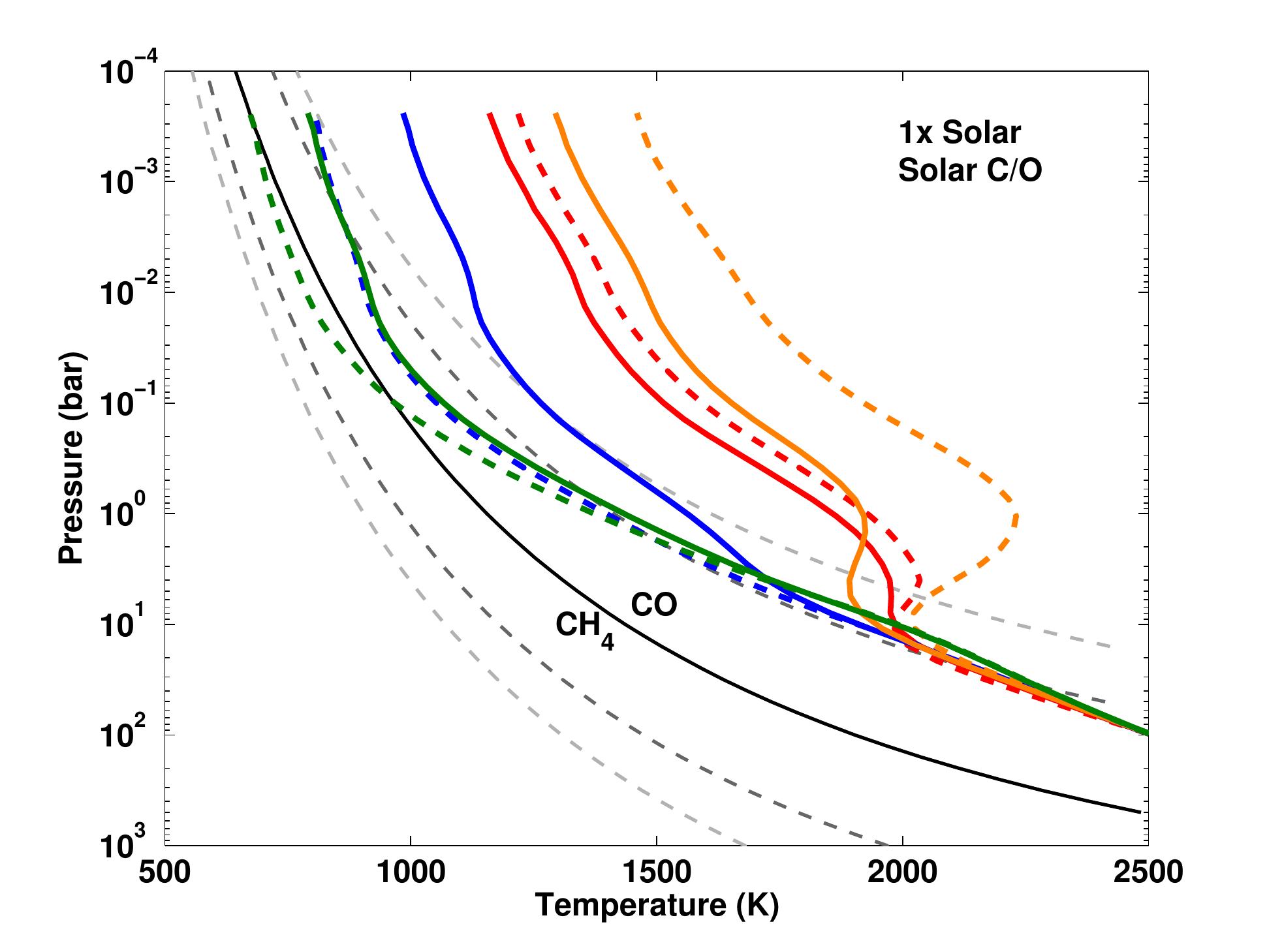}
 \includegraphics[width=0.45\textwidth]{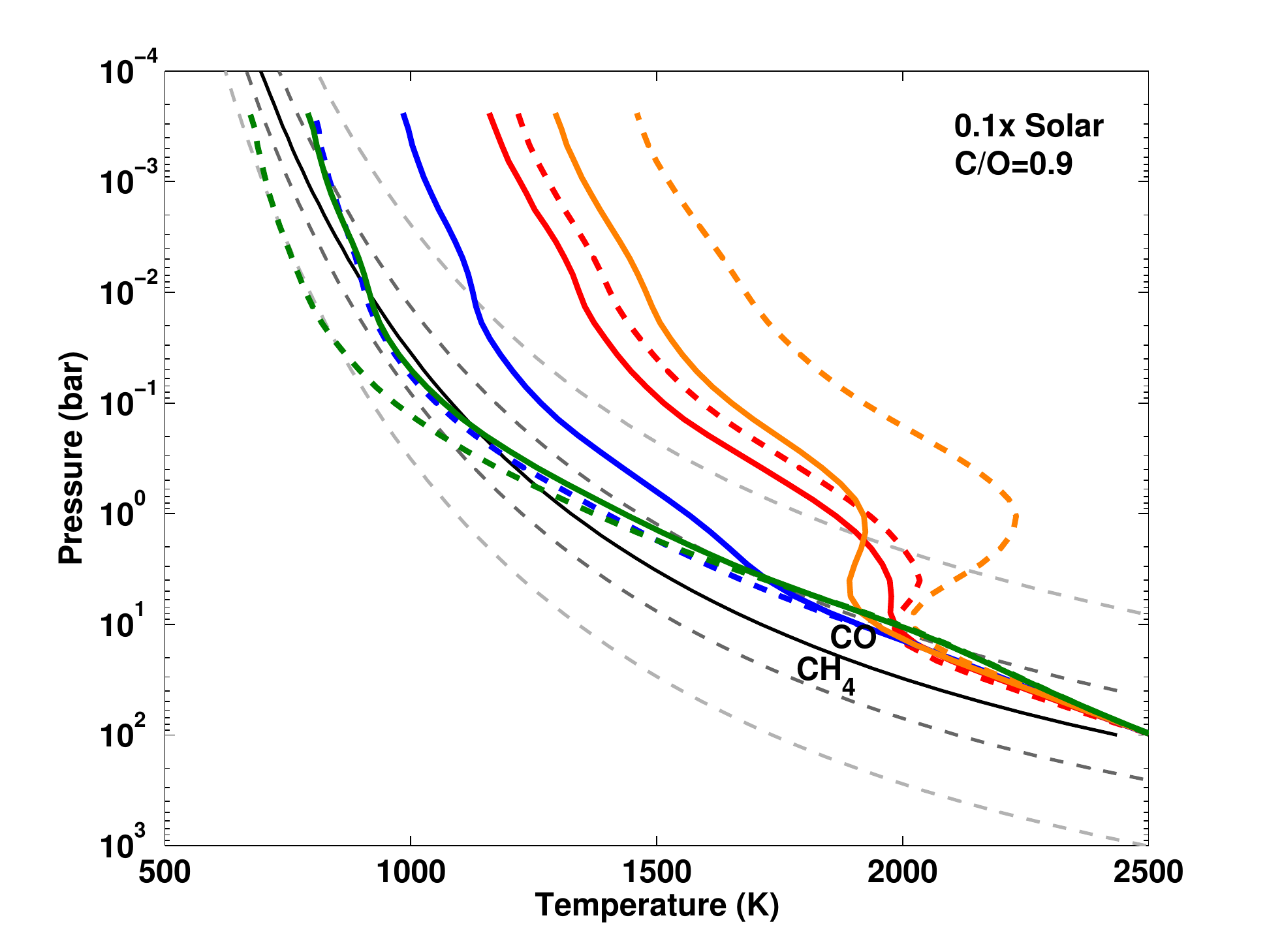}\\
 \includegraphics[width=0.45\textwidth]{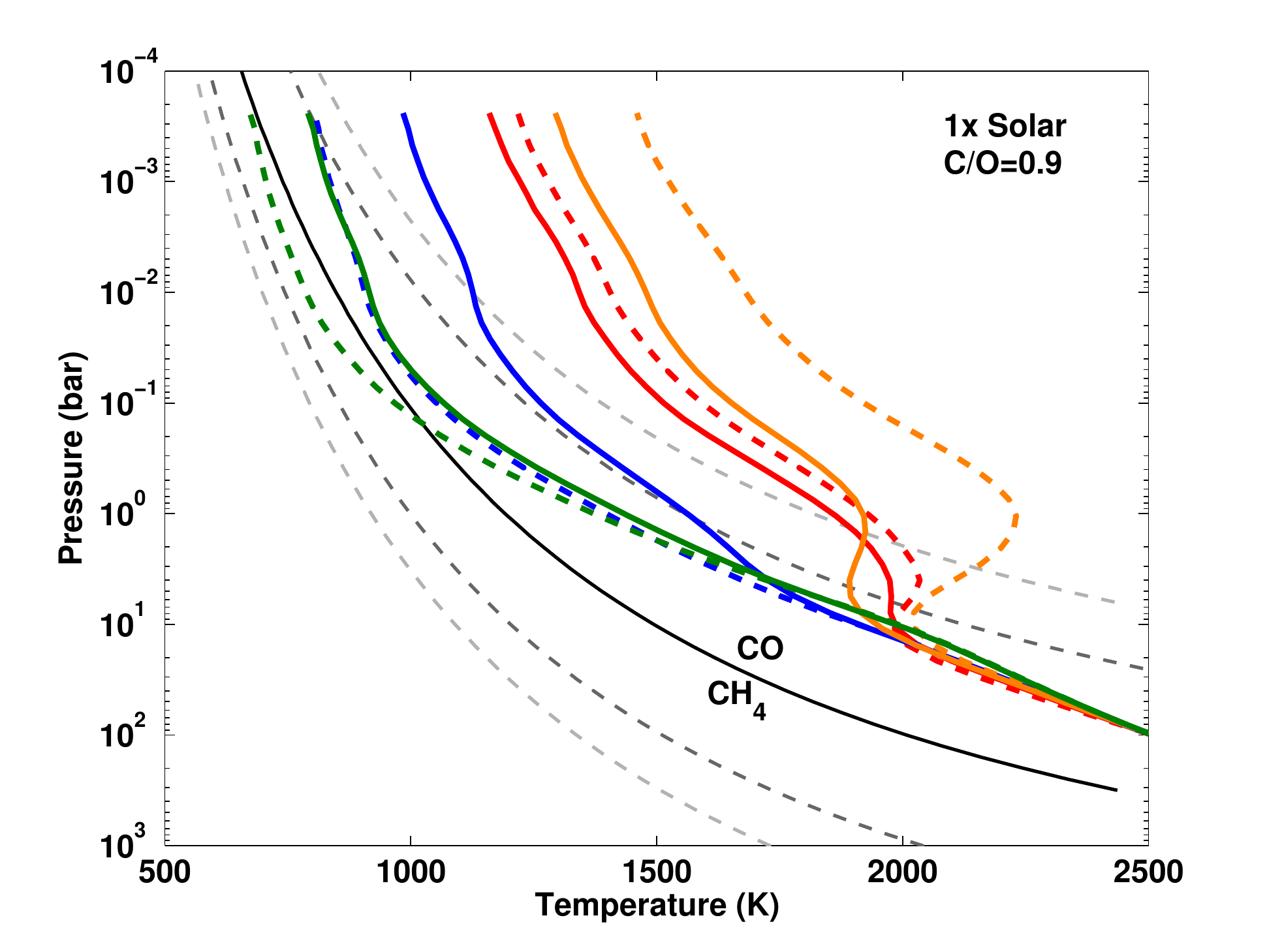}
 \includegraphics[width=0.45\textwidth]{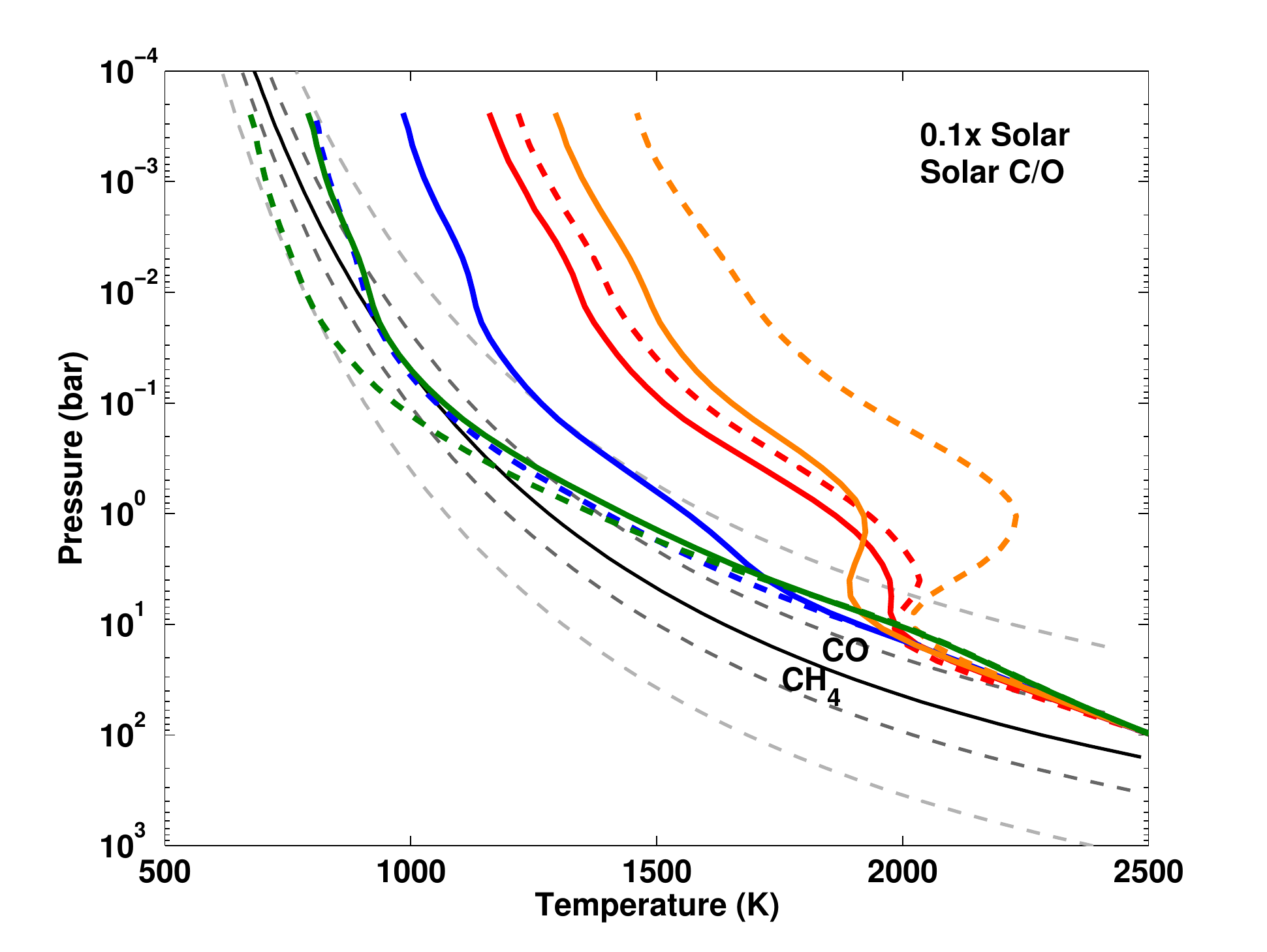}\\
  \caption{Globally averaged (solid lines) and earth-facing hemisphere averaged (dashed lines) pressure temperature profiles from our atmospheric model taken near 
         secondary eclipse (red), periapse (orange), transit (blue), and apoapse (green) compared with CO and CH$_4$ abundance contours assuming solar and 0.1$\times$ solar metallicity 
         as well as C/O=0.55 (solar) and C/O=0.9 atmospheric compositions.  In each panel the solid black line represent the CO=CH$_4$ contour.  The dashed dark grey contours 
         represent the 10 ppm CO (to the left of the solid line) and 10 ppm CH$_4$ (to the right of the solid line) contours.  The dashed light grey contours 
         represent the 1 ppm CO (to the left of the solid line) and 1 ppm CH$_4$ (to the right of the solid line) contours.}\label{hat2_pt_co_ch4}
\end{figure*}

What is perhaps most striking when comparing our model light curves with the observed light curves (Figure~\ref{hat2_model_lc}) 
is that during certain orbital phases our atmospheric models significantly underestimate or overestimate the planetary flux.  
Our models tend to overestimate the planetary 4.5~$\mu$m flux near periapse (in the cases with TiO/VO) and 
underestimate the planetary flux for the bulk of the planetary orbit outside of periapse.  At 3.6~$\mu$m, our atmospheric models 
show deviations from the observations in the opposite sense from what was observed at 4.5~$\mu$m, with the planetary flux 
slightly underestimated near periapse and overestimated near apoapse.   
Especially near apoapse where the planet cools, this discrepancy between our observations and our atmospheric models 
could suggest an enhancement of CH$_4$ and a depletion 
of CO compared with predicted solar metallicity equilibrium abundances.  
If the mixing ratio of CH$_4$ is increased in HAT-P-2b's atmosphere
near apoapse, the amount
infrared flux emitted by the planet at 3.6~$\mu$m will be greatly
reduced, since CH$_4$ is a strong absorber in this bandpass.  
Similarly, a reduction in the mixing ratio of CO allows observations to 
probe deeper at 4.5~$\mu$m, down to generally higher temperatures, leading to an increase in flux.

We now consider how an enhancement of CH$_4$ and depletion of CO in HAT-P-2b's atmosphere, as compared to solar metallicity equilibrium values, 
might occur near the apoapse of its orbit.
Work by \citet{vis12} explores CO/CH$_4$ interconversion timescales for both HAT-P-2b and CoRoT-10b using one-dimensional models.  
In our simulations we find an average vertical windspeed\footnote{\citep[see][for more details on the 
determination of vertical wind speeds from the SPARC model]{lew10}} near the 1~bar level of $\sim$0.6~m~s$^{-1}$.  To get a rough estimate of 
the vertical eddy diffusion coefficient, $K_{zz}$, we simply multiply the vertical wind speed by a relevant vertical length scale which we 
assume to be the scale height, $H$.  From our simulations we estimate a value of $H\sim33$~km, which when combined with our estimate 
of the average vertical windspeed translates into $K_{zz}\sim2\times10^8$~cm$^2$~s$^{-1}$ near the 1~bar level.  We note that this simplistic method of 
estimating $K_{zz}$ might overestimate the actual atmospheric diffusivity as was shown to be the case in simulations of HD~209458b in \citet{par13}.  
Our estimate of $K_{zz}$ places HAT-P-2b in the regime where vertical mixing (as opposed to orbit-induced thermal variations) is 
the dominant quenching mechanism according to \citet{vis12}, which would result in a CH$_4$ mixing ratio of $\sim$10~ppm near the apoapse
of HAT-P-2b's orbit.

One should also consider the possibility of horizontal transport as a possible source of disequilibrium carbon chemistry in HAT-P-2b's atmosphere.  
Such studies have been performed for HD~209458b by \citet{coo06, agu12}.  As can be seen in Figures~\ref{hat2_temp_100mbar_noTiO} and 
\ref{hat2_temp_100mbar_eqchem} a strong eastward equatorial jet is present in our simulations of HAT-P-2b near the apoapse of its orbit.  The 
windspeeds in this zonal jet are $\sim$3500~m~s$^{-1}$ near the 1~bar level.  Assuming that a horizontal dynamical timescale can be estimated as $\tau_{dyn,h}=2\pi R_{p}/U$, 
where $R_p$ is the radius of the planet and $U$ is the zonal windspeed, we estimate $\tau_{dyn,h}\sim10^5$~s.  This horizontal dynamical timescale is significantly shorter than the 
timescale for conversion of CO to CH$_4$ ($\sim10^{10}$~s), and as found by \citet{coo06, agu12} would likely result in a global enhancement of CO.  Overall, it seems 
unlikely disequilibrium chemistry effects, either via horizontal or vertical quenching, are responsible for the discrepancies between our models and the observations of HAT-P-2b.

A significant deviation in the composition of HAT-P-2b's from the assumed solar composition could partially explain the differences 
between the theoretical and observed light curves at 3.6 and 4.5~$\mu$m.  Figure~\ref{hat2_pt_co_ch4} compares globally 
and earth-facing hemisphere averaged pressure-temperature profiles from our simulations near secondary eclipse, transit, periapse, and apoapse 
with CO and CH$_4$ mixing ratios for various atmospheric compositions.  In Figure~\ref{hat2_pt_co_ch4} we have focused on 
cases where the atmospheric metallicity is reduces and/or an enhancement of the C to O ratio compared to solar values is assumed 
since this would naturally cause a depletion in the CO abundance of the planet \citep{mos12, vis12}.  The 
CO mixing ratio could be reduced to as low as 1 ppm at 100~mbar and 10 ppm at 1~bar on the earth-facing hemisphere near apoapse if an atmospheric metallicity of 0.1$\times$ solar 
is considered.  An atmospheric metallicity of 0.1$\times$ solar might seem extreme, but atmospheric chemistries far beyond what is commonly seen in our solar system
have been proposed for a number of planets \citep[e.g.][for GJ~436b]{mos13}.

\section{Conclusions}\label{hat2_conclusions}

We present a three-dimensional atmospheric circulation model 
for HAT-P-2b that incorporates realistic radiative transfer and equilibrium chemical processes
for a range of atmospheric compositions and assumed rotation rates.    
Our atmospheric models reveal the complex radiative and dynamic processes that shape the 
phase curve of HAT-P-2b as observed at 3.6, 4.5, and 8.0~$\mu$m by \citet{lew13}.  Although 
the 3.6 and 8.0~$\mu$m observations strongly favor models that include a thermal inversion near the periapse of 
HAT-P-2b's orbit, the exact timing of the peak in the planetary flux at 3.6, 4.5, and 8.0~$\mu$m 
cannot be explained by a single atmospheric model.  It is likely that processes such as disequilibrium 
chemistry or pressure dependent drag effects, or an atmospheric composition that deviates significantly 
from solar abundances, could further align our predicted phase curves 
with those we observed in \citet{lew13}.
We also find the observed planetary fluxes near apoapse at 3.6 
and 4.5~$\mu$m deviate from our model predictions, which suggest a 
possible variation in CO/CH$_4$ ratio from solar values.

Further work regarding the evolution of HAT-P-2b's atmospheric chemistry throughout its orbit 
is needed to fully explain our observations of HAT-P-2b.  We plan to test how variations in the C/O
ratio as well as a possible sub-solar metallicity composition for HAT-P-2b's atmosphere 
would affect global circulation patterns and theoretical phase variations.
Additionally, we plan to incorporate active tracers in future simulations 
to track CO/CH$_4$ interconversion rates as well as the advection of 
stratosphere-causing particles/species such as TiO.
Initial steps for the incorporation of chemical tracers 
in the SPARC model were taken by \citet{par13} using passive tracers for the case of TiO in the atmosphere of HD~209458b, which is on a 
circular orbit.  Further transit, eclipse, and phase measurements of 
the HAT-P-2 system over a greater wavelength range would help to better constrain the chemistry of 
HAT-P-2b and provide insights into the nature of the dayside inversion as well as the carbon chemistry.  

One key result from this study is that both observations and atmospheric models for HAT-P-2b 
support a short radiative timescale ($\sim$2~hours) near the infrared photosphere of HAT-P-2b 
near periapse.  This measure of the radiative timescale can be used to help inform atmospheric 
models of planets in circular orbits in a similar temperature range, such as HD~209458b.  
Further observations of the HAT-P-2 system at 3.6 and 4.5~$\mu$m have been obtained using 
{\it Spitzer}.  These observations will help us to further refine the exact timing in the peak of the planetary 
flux and allow us to create a two-dimensional dayside map of the planet at 4.5~$\mu$m.
As we expand beyond the realm of close-in hot Jupiters into more earth-like 
worlds questions regarding the habitability of planets on eccentric orbits will certainly arise.  By 
refining our understanding of exoplanets like HAT-P-2b we will be able to use that knowledge 
to constrain the possible circulation, thermal, and chemical processes at work in other exoplanet 
atmospheres.

\acknowledgments

This work was performed in part under contract with the California Institute of Technology (Caltech) funded 
by NASA through the Sagan Fellowship Program executed by the NASA Exoplanet Science Institute.  
APS was supported by NASA Origins grant NNX12AI79G. 
MSM acknowledges support from the NASA PATM program.   
NKL wishes to thank C. Visscher for valuable discussions 
regarding possible chemical processes at work in HAT-P-2b's atmosphere.  The authors thank the 
anonymous referee for their valuable comments on the manuscript.

\bibliographystyle{apj}
\bibliography{ms}

\end{document}